\documentclass[%
reprint,
superscriptaddress,
nofootinbib,
amsmath,amssymb,
aps,
prd,
]{revtex4-2}
\usepackage{graphicx}
\usepackage{dcolumn}
\usepackage{bm}
\usepackage{orcidlink}

\usepackage{physics}
\usepackage[normalem]{ulem} 
\usepackage{aas_macros}
\usepackage{subfigure}

\usepackage{booktabs} 
\usepackage{siunitx}  
\usepackage{enumerate}

\definecolor{RED}{rgb}{1,0,0} 
\usepackage{ragged2e} 
\usepackage[utf8]{inputenc}

\usepackage{multirow} 
\usepackage{placeins}
\usepackage{hyperref}
\usepackage[all]{hypcap}

\newcommand\rk{\color{purple}}

\usepackage[T1]{fontenc}
\begin{document}

\title{Little Red Dots as Hidden Neutrino Sources}

\author{Riku Kuze \orcidlink{0000-0002-5916-788X}}
\email{riku.kuze@yukawa.kyoto-u.ac.jp} 
\affiliation{Center for Gravitational Physics and Quantum Information,
Yukawa Institute for Theoretical Physics, Kyoto, Kyoto 606-8502 Japan}

\author{Kunihito Ioka \orcidlink{0000-0002-3517-1956}}
\email{kunihito.ioka@yukawa.kyoto-u.ac.jp}  
\affiliation{Center for Gravitational Physics and Quantum Information, Yukawa Institute for Theoretical Physics, Kyoto, Kyoto 606-8502 Japan}

\author{Kohta Murase \orcidlink{0000-0002-5358-5642}}
\email{murase@psu.edu}  
\affiliation{Department of Physics; Department of Astronomy \& Astrophysics; Center for Multimessenger Astrophysics, Institute for Gravitation and the cosmos, The Pennsylvania State University, University Park, Pennsylvania 16802, USA}
\affiliation{Center for Gravitational Physics and Quantum Information,
Yukawa Institute for Theoretical Physics, Kyoto, Kyoto 606-8502 Japan}

\author{Shigeo S. Kimura \orcidlink{0000-0003-2579-7266}}
\email{shigeo@astr.tohoku.ac.jp}  
\affiliation{Frontier Research Institute for Interdisciplinary Sciences, Tohoku University, Sendai 980-8578, Japan}
\affiliation{Astronomical Institute, Graduate School of Science, Tohoku University, Sendai 980-8578, Japan}

\author{Kohei Inayoshi \orcidlink{0000-0001-9840-4959}}
\email{inayoshi.pku@gmail.com}
\affiliation{Kavli Institute for Astronomy and Astrophysics, Peking University, Beijing 100871, China}

\date{\today}

\begin{abstract}
Little Red Dots (LRDs) are enigmatic, compact, red galaxies at high redshift, $z\sim 4$--$7$, discovered by the James Webb Space Telescope. Broad emission lines in the absence of X-ray and radio counterparts suggest that they host accreting supermassive black holes embedded in dense gaseous envelopes. This black-hole-envelope configuration facilitates efficient photohadronic interactions and neutrino production. Remarkably, their observed source number density and luminosity are compatible with the energetics of the diffuse neutrino background. We consider that relativistic jets and outflows are launched from the black hole and propagate through low-density polar funnels within envelopes, where particle acceleration and neutrino emission occur. This leads to LRDs being effectively hidden sources. Our analytic and numerical calculations show that, in an optimistic scenario, LRDs can contribute $\sim 30\%$ of the observed diffuse background at TeV--sub-PeV energies, predominantly through photomeson production. At high neutrino energies, $\gtrsim 10^{5.5}~{\rm GeV}$, inverse-Compton cooling of muons modifies the resulting flavor ratio, providing a distinctive diagnostic for IceCube--Gen2 and other upcoming neutrino telescopes.
\end{abstract}

\maketitle


\section{Introduction}\label{sec:intro}
Recently, the James Webb Space Telescope (JWST) has revealed a population of compact and red galaxies at high redshifts, called Little Red Dots (LRDs) \citep{2023ApJ...959...39H, Kocevski2023ApJ, Matthee2024ApJ, Kocevski2024arXiv, Akins2025, Labbe2025ApJ}. They have been detected at $z \simeq 4$–7 with an inferred comoving number density of $n \simeq 10^{-5}$–$10^{-4}~\rm cMpc^{-3}$ at $z\simeq5$ \citep{Kocevski2023ApJ, Akins2025}, orders of magnitude higher than that of luminous quasars known at similar epochs \citep{McGreer2018AJ, Matsuoka2018ApJ, Niida2020ApJ, Matsuoka2023ApJ}. 

Several observational properties of LRDs suggest that they host accreting supermassive black holes (SMBHs; see \citealt{Inayoshi_Ho2026}, references therein). The detection of broad Balmer emission lines, with widths of several thousands km~s$^{-1}$, points to gas bound to a central massive object, analogous to the broad-line regions of active galactic nuclei (AGNs) \citep{Kocevski2023ApJ, Matthee2024ApJ, Greene+:2024}. In addition, after correcting for dust obscuration, the inferred bolometric luminosities reach $L_{\rm bol, inf}\simeq10^{45-47}$~erg~s$^{-1}$, while no detections of radio and X-ray signals have been reported \citep{Kocevski2023ApJ, Yue2024ApJ, Ananna2024ApJ, Akins2025}. The observed $L_{\rm bol, obs}\simeq10^{44-45}$~erg~s$^{-1}$ is comparable to the Eddington luminosity for a black hole (BH) mass of $M_{\rm BH}\simeq10^{6-7}M_\odot$, where $L_{\rm Edd}\simeq1.3\times10^{38}(M_{\rm BH}/M_\odot)$~erg~s$^{-1}$, and this mass range is consistent with that independently estimated from the empirical relations established for classical AGN \citep[e.g.,][]{Matthee2024ApJ,2024ApJ...974..147Lin, 2025ApJ...986..165Taylor, 2026ApJ...996...93Lin}. This implies that the accretion rate is comparable to or exceeds the Eddington rate. In such a near- or super-Eddington regime, the inner accretion flow is Compton thick, which naturally accounts for the current non-detections in the radio and X-ray bands. However, super-Eddington accretion flows are expected to produce powerful radiation-driven outflows \citep{ss73, BB99a, Blandford+B:2004, OMN05a, OM11a, Jiang:2014wga, Sadowski+NMT:2014, Sadowski+:2015, Inayoshi2016MNRAS}, which imprint strong feedback on the ambient medium. The associated emission should also be present.

To resolve this discrepancy, \citet{2025MNRASKido} proposed that the central SMBH embedded in a dense, optically thick envelope may explain LRDs. In this scenario, the intrinsic emission from the accretion flow is trapped and thermalized within the envelope, and only emerges at the photosphere as thermal optical/infrared radiation, giving rise to the characteristic red optical and flat infrared spectral shape while suppressing direct X-ray signatures \citep{2025arXiv250505322IN}. Such an environment has important implications for high-energy processes. A similar idea of a BH embedded in an envelope had been discussed in 1970s in the context of neutrino production~\citep[e.g.,][]{Berezinsky1977,1981MNRAS.194....3B}. 

IceCube detected the diffuse neutrino background in the TeV--PeV range, yet its astrophysical origin remains a major open problem \citep{2016ApJ...833....3A, Aartsen:2018vtx, 2019ICRC...36.1017S, IceCube2024PRD, IceCube2025A}. Hadronic interactions that produce neutrinos generate high-energy gamma rays. If all the sources that contribute to the diffuse neutrino background were gamma-ray transparent, the resulting sub-TeV--10 TeV gamma rays would exceed the observed gamma-ray background. This tension requires that a substantial fraction of the diffuse neutrino background originates from gamma-ray–opaque, ``hidden'' neutrino sources \citep{Murase:2015xka,2016ApJ...826..133X,2017ApJ...836...47B} \footnote{Cosmological attenuation of TeV gamma rays by the extragalactic background light is insufficient to avoid the gamma-ray constraint on the origin of neutrinos at least in $pp$ scenarios. As investigated in \citet{2016ApJ...826..133X}, if high-redshift sources dominate the diffuse neutrino background, the produced gamma rays cascade into the 10~GeV band, and the resulting emission can approach the Fermi-LAT extragalactic gamma-ray background (see Fig.~6 of \cite{2016ApJ...826..133X}). A substantial fraction of the 10~GeV gamma-ray background is still attributed to blazars \citep{2015ApJ...800L..27A}. Thus, a dominant contributor to the diffuse neutrino background requires suppressed gamma-ray escape at the source, even though the constraint is somewhat alleviated for high-redshift sources.}. Among many source classes that have been proposed \citep[e.g.,][]{Murase:2016gly}, this condition is naturally realized in the vicinity of SMBHs, and the candidate sources include AGNs \citep[e.g.,][]{2023ecnp.book..483M}, tidal disruption events \citep[e.g.,][]{2017ApJ...838....3S, Murase:2020lnu, Mukhopadhyay:2023mld}, and LRDs since their dense environments prevent the escape of gamma rays from the system. Because LRDs are mainly distributed at high redshifts, the detection of neutrinos from individual LRDs would be challenging. However, their large number density may allow a sizable cumulative contribution to the diffuse neutrino background. High-redshift sources such as Pop-III gamma-ray bursts and Pop-III supernovae have also been proposed as potential contributors \citep{2002MNRAS.334..173S, Iocco:2007td, Berezinsky2012PhRvD, 2016ApJ...826..133X}.

In this work, we propose that LRDs constitute a previously unrecognized class of high-redshift neutrino sources, as their BH-envelope systems provide conditions for efficient neutrino production and their large comoving number density suggests a potentially significant contribution to the diffuse neutrino background. In this framework, we consider that the jet dissipates within the envelope \citep{Czerny+2012ApJL}, and that the resulting non-thermal protons interact with photons from the surrounding disk, producing neutrinos. Section \ref{sec:diffuse_convert} gives a model-independent, order-of-magnitude estimate of the potential LRD contribution to the diffuse neutrino background, before turning to detailed emission models. In Section \ref{sec:funnel}, we outline the basic structure of the BH-envelope system and examine the conditions for particle acceleration within it. In Section \ref{sec:neutrino}, we evaluate the maximum cosmic-ray energy and the resulting neutrino spectrum, both analytically and with the Astrophysical Multi-messenger Emission Simulator (AMES) code \citep{2023MNRAS.524...76Z, 2025arXiv251223231Wei}. We evaluate the collective contribution of LRDs to the diffuse neutrino background in Section \ref{sec:diffuse_neu}. In Section \ref{sec:discussion}, we discuss the detectability of individual LRDs, examine observational tests to distinguish them from radio-quiet AGNs, and compare their physical environment with that of conventional AGN jets. We conclude in Section \ref{sec:conclusion}.

For clarity, throughout this paper, we adopt the notation $Q_x \equiv Q/10^x$ in cgs units and normalize masses to the solar mass, $M_\odot$. Unless otherwise noted, we assume a flat $\Lambda$CDM cosmology with $H_0 = 70~\rm km~s^{-1}~Mpc^{-1}$, $\Omega_m = 0.3$, and $\Omega_\Lambda = 0.7$. In what follows, primed quantities ($^\prime$) are measured in the jet comoving frame, and unprimed quantities are defined in the source rest frame at the emitter redshift. We denote the particle energy in the comoving frame by $\varepsilon'$, that in the source rest frame by $\varepsilon$, and that in the observer frame at $z=0$ by $E$. We also use the subscript “$\nu$” to denote neutrino-related quantities, such as $L_\nu$ for neutrino luminosity ($\rm erg~s^{-1}$) and $E_\nu$ for neutrino energy ($\rm erg$).

\section{Population-luminosity diagram for LRDs}\label{sec:diffuse_convert}
In this section, we first provide order-of-magnitude estimates of the effective neutrino luminosity and the comoving source density of LRDs. We examine whether these quantities are sufficient to account for the diffuse neutrino background. Many source classes are abundant in the local Universe, and their diffuse contributions are commonly evaluated by using their effective local number density, $n_{0}^{\rm eff}$. However, LRDs are absent in the local Universe, making $n_{0}^{\rm eff}$ an inadequate quantity for this evaluation. For such high-redshift-dominant populations, we rescale the source density as $\xi_z n_{0}^{\rm eff}$, where $\xi_z$ is a factor reflecting the population evolution. The explicit form of $\xi_z$ is defined later. Using this prescription with typical LRD parameters, we discuss the implications within the framework of order-of-magnitude estimates.

We begin with simple estimates of the neutrino luminosity and the comoving number density of a typical LRD. As a representative site for cosmic-ray acceleration, we consider a relativistic jet launched from a central BH with $M_{\rm BH}\simeq10^{6.5}M_\odot$ and approximate its power by the Eddington luminosity, $L_j\approx L_{\rm Edd}$. This treatment is equally applicable to mildly relativistic outflows, which may be more readily realized in such environments. We assume that a fraction $\epsilon_p \simeq0.1$ of this power is carried by accelerated (non-thermal) protons {\rk \citep[e.g.,][]{CS2009, SSA13a, CS14a}}. Given the copious photons and dense material in the envelope, accelerated protons efficiently produce neutrinos before escape, i.e., the pion-production efficiency is almost $100\%$. Then, the corresponding per-flavor neutrino luminosity is approximated to be
\begin{eqnarray} 
L_{\nu_\mu} &\approx& \frac{1}{8}\epsilon_p f_{\rm bol}L_j \nonumber \\ 
&\simeq& 5.0\times10^{41}~{\rm erg~s^{-1}}~\epsilon_{p,-1}f_{\rm bol, -1}L_{j,44.6} \label{eq:Lnu_order}
\end{eqnarray} 
where $f_{\rm bol}\simeq0.1$ is the bolometric correction factor introduced to estimate the neutrino luminosity around some energy range. The factor $1/8$ reflects a simple energy partition argument. In a photon-rich environment, most of the proton energy is lost through photomeson production ($p+\gamma \rightarrow p+\pi$). About half of the resulting pion energy is carried by charged pions, whose decay energy is shared among the three neutrino flavors and one electron/positron, yielding about one-eighth per flavor. This neutrino luminosity is only a rough estimate, and more detailed modeling is presented in this work.

Observationally, the redshift evolution of LRDs is suggested to follow a lognormal form \citep{2025ApJ...988L..22I,2025arXiv250800057T}, and we adopt this prescription in the present estimates (see Appendix~\ref{app:numberdensity}, \textit{Lognormal model}, for the detailed expression). We set the normalization so that the comoving source number density, $n(z)$, peaks at $5\times10^{-5}~{\rm cMpc^{-3}}$ around $z\simeq5$--$7$ \citep{2023ApJ...959...39H, Kocevski2024arXiv, Kokorev+:2024, 2025ApJ...989L..50C}.

Population–luminosity diagrams are widely used to discuss whether specific source classes can account for the diffuse neutrino background \citep{Murase:2016gly, Aartsen:2019Gen2}. We place LRDs on the same plot, but with rescaled quantities. In the original formulation \citep{2014PhRvD..90d3005Ahlers,Murase:2016gly}, source classes are represented by their local ($z\simeq0$) density, and redshift evolution is taken into account through the diffuse and multiplet curves. This representation is not suitable for LRDs, whose number density approaches zero at $z=0$. To make high-redshift-dominated populations comparable on the same plane, we rescale the vertical axis to $\xi_z n_{0}^{\rm eff}$. This rescaling enables their diffuse and multiplet implications to be compared directly with those of other source classes on the same diagram.

First, we derive the diffuse--requirement line on the $(L_{\nu,{\rm eff}},\xi_z n_{0}^{\rm eff})$ plane. Following the standard population--luminosity formulation \citep[e.g.,][]{mal13,2020ApJ...890...25Y}, the diffuse neutrino intensity can be written as
\begin{align}
E_\nu^2\Phi_\nu \approx \frac{ct_H}{4\pi}L_{\nu,{\rm eff}} n_0^{{\rm eff}}\xi_z,
\label{eq:dif_base}
\end{align}
where $\Phi_\nu$ is the energy-differential diffuse neutrino intensity, $t_H$ is the Hubble time, $L_{\nu,{\rm eff}}$ is the per-source and per-flavor neutrino luminosity in the source rest frame, and $c$ is the speed of light. Following the form of Eq.~(5) in \citet{WB99a}, the evolution factor $\xi_z$ is given by
\begin{align}
\xi_z = \frac{\int^{z_{\max}}_0 dz (1+z)^{-1} \left|\frac{dt}{dz}\right| \frac{n(z)}{n_0^{\rm eff}}}{\int^\infty_0 dz \left|\frac{dt}{dz}\right|},
\label{eq:xi_z}
\end{align}
where $dt/dz = -1/((1+z)H(z))$ is the cosmic time-redshift relation and $H(z) = H_0E(z)$ is the Hubble parameter at redshift $z$ with $E(z) = \sqrt{\Omega_m(1+z)^3 + \Omega_\Lambda}$. We adopt $z_{\max}\simeq 10$ as a maximum redshift, which is consistent with the recent report of the high-redshift LRD candidates \citep{2025arXiv250800057T}. From Eq.~(\ref{eq:dif_base}) and the observed diffuse neutrino background \citep[see Eq.~(6) of ][]{Murase:2016gly}, a population can account for the observed diffuse neutrino background if
\begin{align}
\xi_z n_0^{\rm eff} \simeq 4.8\times10^{-7}~{\rm Mpc^{-3}} &\left(\frac{E_\nu^2\Phi_\nu}{10^{-8}~{\rm GeV~cm^{-2}~s^{-1}~sr^{-1}}}\right) \nonumber \\
&\times L_{\nu,\rm eff, 42}^{-1}.
\label{eq:xiz_n_diff}
\end{align}
Since the redshift evolution is incorporated into $\xi_z$, the diffuse requirement line takes the form of a single straight line in the $(L_{\nu,{\rm eff}},\xi_z n_0^{\rm eff})$ plane, independent of the specific redshift distribution of each source class.

We next rewrite the multiplet constraint. Following the method of \citet{Murase:2016gly}, the expected number of sources yielding at least two events is
\begin{equation}
N_{m\ge 2}(L_{\nu, \rm eff})=\Delta\Omega \int_0^{z_{\max}} \frac{dV_c}{dzd\Omega} P_{m\ge2} \big(s\big) n(z) dz,
\label{eq:multiplet}
\end{equation}
where $\Delta\Omega\simeq 2\pi$ is the surveyed solid angle and $dV_c/(dzd\Omega)= d_c^2(z) (c/H(z))$ is the comoving volume element with $d_c(z)=\int_0^z c/{H(z')} dz'$ being the comoving distance. The Poisson probability for obtaining at least two events is $P_{m\ge2}(s)=1-e^{-s}(1+s)$, where $s=(d_{N=1}(L_{\nu, \rm eff},F_{\lim})/d_L(z))^2$ is the count and $d_L(z)=(1+z)d_c(z)$ is the luminosity distance. The distance $d_{N=1}(L_{\nu,\rm eff},F_{\lim}) \approx (L_{\nu, \rm eff}/(4\pi F_{\rm lim}/2.4))^{1/2} \simeq 110~{\rm Mpc} ~ L_{\nu, \rm eff, 42}^{1/2}F_{\rm lim, -9}^{-1/2}$ is that one event is expected from the sources, where $F_{\rm lim, -9}=F_{\rm lim}/(10^{-9}~{\rm GeV~cm^{-2}~s^{-1}})$ is the IceCube 90\% C.L. upper limit \citep[see Eq.~(4) of][]{Murase:2016gly}. At high energies, the IceCube data are effectively background-free, and the 90\% C.L. upper limit on the Poisson mean for zero observed events is 2.44, hence the single-event-equivalent flux is approximated to be $F_{\lim}/2.4$ \citep{1998PhRvDFeldmanCousins}.

For a convenient representation of the multiplet expression, \citet{Murase:2016gly} introduced a luminosity-dependent function $q_L$ that accounts for the redshift-evolution integral. In their formulation, the expected number of sources producing $m\ge2$ multiplets is written as
\begin{equation}
N_{m\ge 2}(L_{\nu,\rm eff}) = \sqrt{\pi}q_L \left(\frac{\Delta\Omega}{3}\right) n_{0}^{\rm eff} d_{N=1}^{3},
\label{eq:MW16-3}
\end{equation}
which is equivalent to Eq.~(\ref{eq:multiplet}). By equating Eq.~(\ref{eq:multiplet}) and Eq.~(\ref{eq:MW16-3}), $q_L$ is given by
\begin{equation}
q_L= \frac{3 \int_0^{z_{\max}} \frac{dV_c}{dzd\Omega} P_{m\ge2} \big(s\big) \frac{n(z)}{n_0^{\rm eff}} dz}{\sqrt{\pi}d_{N=1}^{3}}. 
\label{eq:qL}
\end{equation}
The critical condition for non-detection is $N_{m\ge2}<1$, which yields the multiplet--limit condition approximately given in Eq.~(5) of \citet{Murase:2016gly}. To express this constraint on the same vertical axis used for the diffuse requirement, we multiply both sides of Eq.~(5) of \citet{Murase:2016gly} by $\xi_z$, obtaining
\begin{equation}
\xi_z n_{0}^{\rm eff}\lesssim 1.9\times10^{-7}~{\rm Mpc^{-3}}\left(\frac{\xi_z}{q_L}\right)\left(\frac{2\pi}{\Delta\Omega}\right) L_{\nu,{\rm eff}, 42}^{3/2} F_{\lim,-9}^{3/2}.
\label{eq:xi-ncrit}
\end{equation}
This expression provides an approximate representation of the multiplet--limit curves in the $(L_{\nu,{\rm eff}},\xi_z n_{0}^{\rm eff})$ plane. Because both $\xi_z$ and $q_L$ depend on the redshift evolution of each population, different source classes yield different multiplet--limit curves. 

\begin{figure}[t]
    \centering
    \includegraphics[width=\linewidth]{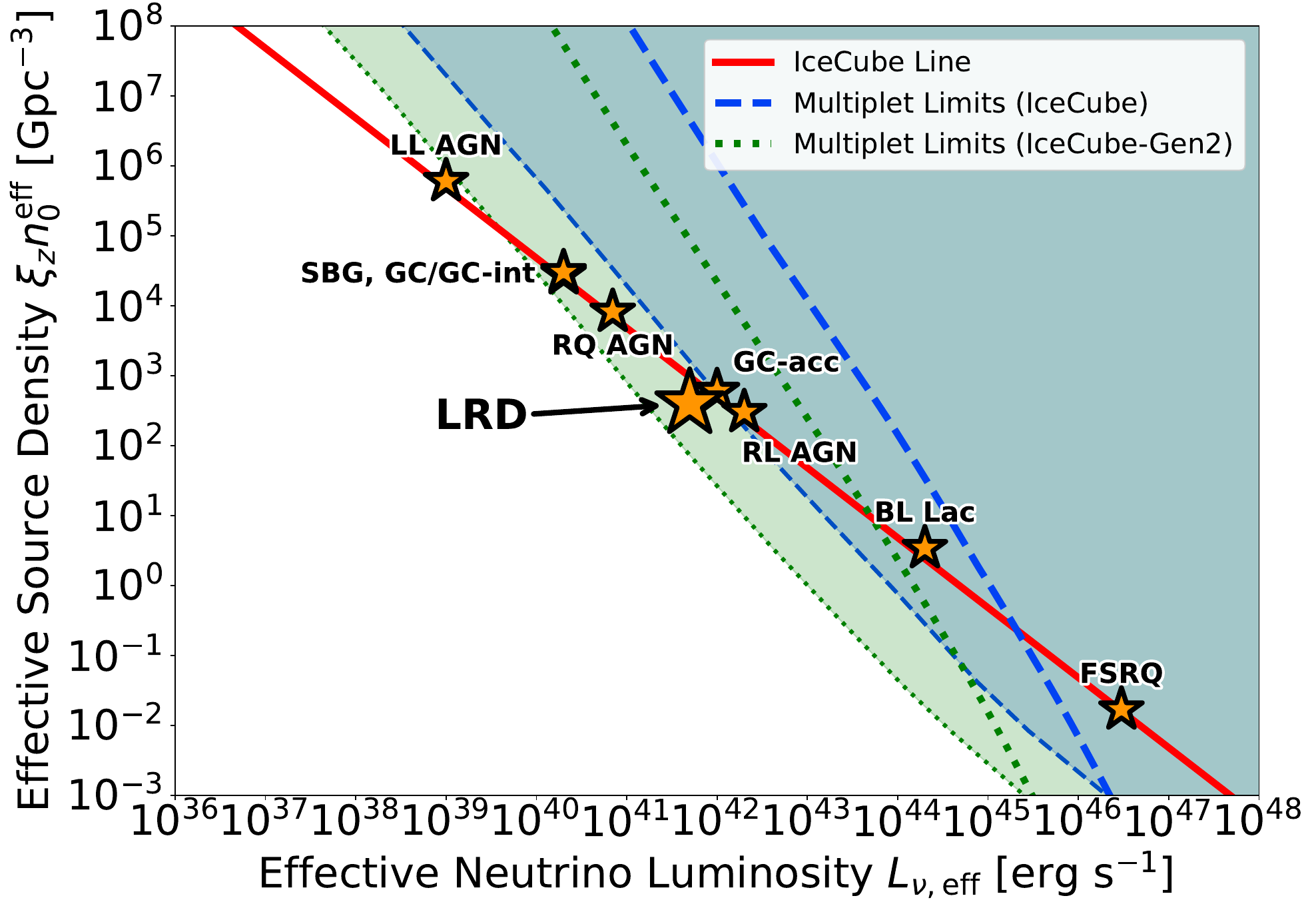}
    \caption{Rescaled population--luminosity diagram in the $(L_{\nu,{\rm eff}},\,\xi_z n_0^{\rm eff})$ plane. Orange-filled stars mark the values of $(L_{\nu,{\rm eff}}, \xi_z n_0^{\rm eff})$. The red solid curve denotes the diffuse requirement (Eq.~(\ref{eq:xiz_n_diff})).  The multiplet--limit curves are shown for IceCube with six-year data as blue-dashed line \citep{2015arXiv151005222T} and for IceCube--Gen2 as green-dotted line \citep{Gen2_2021}. The thin multiplet--limit curves show the multiplet limits based on the SFR evolution, taken from \citet{Murase:2016gly}, rescaled by the SFR evolution factor $\xi_{z,\rm SFR} \simeq 2.8$. The thick curves represent the multiplet limits after converting the SFR-based result to the LRD case by using the LRD-specific evolution factor $\xi_{z,\rm LRD}$ (Eq.~(\ref{eq:xi_z})) and luminosity-dependent function $q_{L, \rm LRD}$ (Eq.~(\ref{eq:qL})). For the LRD point, we use $L_{\nu,{\rm eff}}\simeq5.0\times10^{41}~{\rm erg~s^{-1}}$ (Eq.~(\ref{eq:Lnu_order})) and the lognormal form of $n(z)$ described in Appendix~\ref{app:numberdensity}. For comparison, we also show typical locations of FSRQs, BL~Lacs, starburst galaxies (SBGs), galaxy clusters or groups (GC/GG--int), radio-loud AGN (RL~AGN), radio-quiet AGN (RQ~AGN), and low-luminosity AGN (LL~AGN) \citep{Murase:2016gly}.}
    \label{fig:rho-Lnu}
\end{figure}

Fig.~\ref{fig:rho-Lnu} presents a population–luminosity diagram with $(L_{\nu, \rm eff}, \xi_z n_0^{\rm eff})$. The overlaid curves (diffuse requirement and multiplet) represent the observational thresholds (see Eqs.~(\ref{eq:xiz_n_diff}) and (\ref{eq:xi-ncrit})). The diffuse--requirement line, shown as the red solid line, is constructed with $E_\nu^2\Phi_\nu=10^{-8}~{\rm GeV~cm^{-2}~s^{-1}~sr^{-1}}$ \citep{2016ApJ...833....3A, IceCube2024PRD, IceCube2025A}. For the multiplet--limit curves, we use $F_{\rm lim}^{\rm IC}\simeq (6\text{--}7)\times10^{-10}~{\rm GeV~cm^{-2}~s^{-1}}$, shown as the blue-dashed curve, and $F_{\rm lim}^{\rm Gen2} \simeq 1.0\times10^{-10}~{\rm GeV~cm^{-2}~s^{-1}}$, shown as the green-dotted curve. The former corresponds to the IceCube six-year sensitivity \citep{2015arXiv151005222T}, and the latter to the projected IceCube--Gen2 ten-year sensitivity \citep{Gen2_2021}. These choices follow \citet{Murase:2016gly} for simplicity. The thin multiplet--limit curves represent the multiplet limits obtained by adopting the star-formation rate (SFR) evolution and are taken from \citet{Murase:2016gly}, rescaled by the SFR evolution factor $\xi_{z,\rm SFR} \simeq 2.8$. The thick curves show the multiplet limits for the LRD case. To construct these curves, we compute $\xi_z$ (Eq.~(\ref{eq:xi_z})) and $q_L$ (Eq.~(\ref{eq:qL})) for both the SFR and the LRD evolutions. The LRD values of $\xi_z$ and $q_L$ are estimated using $L_{\nu,{\rm eff}}\simeq 5.0\times10^{41}~{\rm erg~s^{-1}}$ from Eq.~(\ref{eq:Lnu_order}) and the lognormal form of $n(z)$ (see Appendix~\ref{app:numberdensity}). For the SFR evolution and luminosity, we adopt the values used in \citet{Murase:2016gly}. The SFR-based multiplet--limit curves are then rescaled by $\xi_{z, \rm LRD}q_{L,\rm SFR}/(\xi_{z, \rm SFR}q_{L,\rm LRD})$ to obtain the LRD multiplet--limit curves shown as thick curves. We plot $L_{\nu,{\rm eff}}$ and $ \xi_z n_0^{\rm eff}$ for each source class as orange-filled stars with black edges. For non--LRD source classes, the adopted redshift evolution and $L_{\nu,{\rm eff}}$ follow Table~1 of \citet{Murase:2016gly}.

The point for LRDs shown in Fig.~\ref{fig:rho-Lnu} lies close to the diffuse--requirement line, indicating that the population may contribute to the observed diffuse neutrino background. This motivates more detailed modeling of neutrino emission from LRDs, developed later. The same figure also shows that the LRD point is far below the multiplet--limit curves, which implies that multiplet detections with IceCube or IceCube--Gen2 are unlikely. The multiplet limits for LRDs lie well above those of other source classes. This arises because LRDs are concentrated at high redshift and nearly absent at $z\simeq0$, yielding a higher $\xi_z$ than for populations residing primarily in the nearby Universe.


\section{Jet in a funnel of a BH envelope} \label{sec:funnel}
In the BH-envelope scenario for LRDs, the dense and optically thick envelope is collisional and not suitable for cosmic-ray acceleration. For efficient acceleration, a low-density region is required. Matter near the rotation axis has little angular momentum and falls almost freely onto the BH, which leaves the polar region tenuous. At larger polar angles, material retains angular momentum and forms an accretion disk. This contrast naturally produces an evacuated polar region, called a funnel. In this work, we consider that the relativistic jet launched from the central BH propagates and dissipates in this funnel, providing the site for particle acceleration. Mildly relativistic outflows can also be described within the same framework. 

Since the accretion proceeds at super-Eddington rates, photons are trapped, and radiation-driven outflows are launched. The outflowing gas is mixed with the envelope and partially falls back onto the accretion flow, inflating the accretion flows and forming a convective disk \citep{2025MNRASKido}. Thus, the accretion flow is expected to be a convection-dominated accretion flow \citep[CDAF;][]{Narayan+IA:2000, 2000ApJ...539..809Q}. We consider a jet launched from the central SMBH in this CDAF background, with $L_j\approx L_{\rm Edd}$. In the fiducial case, the gas and radiation pressures of the surrounding accretion flow exceed the jet ram pressure, meaning that the funnel geometry and dynamics are determined primarily by the CDAF rather than by the jet itself.

\begin{figure}[tb]
    \centering
    \includegraphics[width=\linewidth]{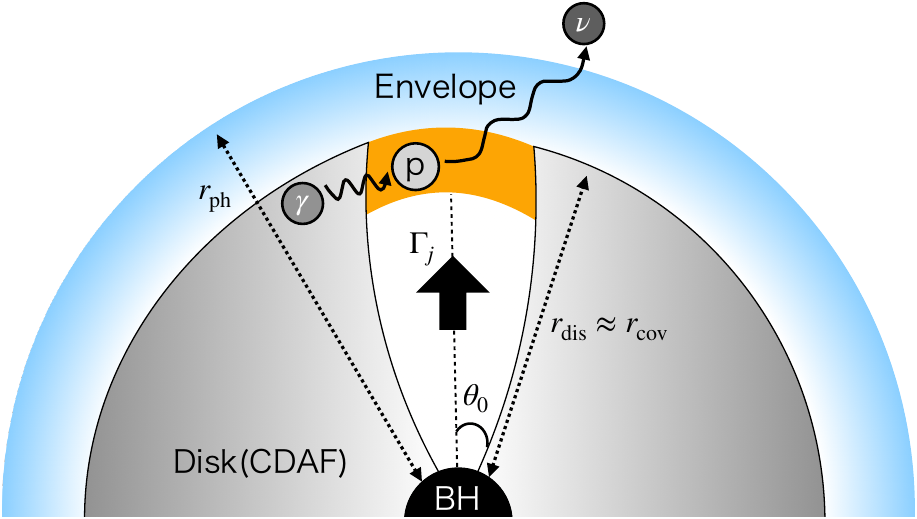}
    \caption{Schematic image of the BH-envelope-jet system considered in this work. A central SMBH is embedded in an envelope that extends outward to the photospheric radius $r_{\rm ph}$. In the inner region, gas with nonzero angular momentum circularizes into an accretion disk, whereas along the rotation axis the low-angular-momentum inflow can fall almost in free fall toward the BH, creating a low-density polar funnel. The funnel extends from the base at $r_0=r_{\min}R_g$ near the BH horizon to the covered radius, $r_{\rm cov}$, beyond which the envelope globally confines the jet. The funnel opening angle at its base is $\theta_0$, and a relativistic jet with Lorentz factor $\Gamma_j$ propagates inside the funnel. The orange-shaded region indicates the jet dissipation region. For simplicity, we set the dissipation radius to be close to the covered radius, $r_{\rm dis}\approx r_{\rm cov}\simeq 10^{16}$ cm. In this region, photons from the surrounding accretion flow provide a dense target for cosmic-ray interactions, leading to high-energy neutrino production. The corresponding comoving baryon number density and the comoving photon number density at the dissipation region are $\sim 4\times10^{7}~\rm cm^{-3}$ and $\sim 2\times10^{17}~\rm cm^{-3}$, respectively (see Eqs.~(\ref{eq:nj_prime}) and (\ref{eq:ph_num_density})). At the same radius, the envelope baryon number density is $n_{\rm env}\sim 3\times10^{13}~\rm cm^{-3}$, as estimated from Eq.~(\ref{eq:rho}).}
    \label{fig:schematic}
\end{figure}

We illustrate the BH-envelope-jet geometry in Fig.~\ref{fig:schematic}. The envelope extends up to photosphere, 
\begin{eqnarray}
r_{\rm ph}
&=&\left(\frac{L_{\rm env}}{4\pi \sigma_{\rm SB} T_{\rm eff}^{4}}\right)^{1/2} \nonumber \\
&\simeq& 3.0\times10^{16}~{\rm cm }~ L_{\rm env,44.6}^{1/2}\left( \frac{T_{\rm eff}}{5000~\rm K}  \right)^{-2},
\label{eq:rph}
\end{eqnarray}
where $L_{\rm env}$ is the envelope luminosity, $T_{\rm eff}$ is the effective temperature, and $\sigma_{\rm SB}$ is the Stefan--Boltzmann constant. Throughout this work, we adopt $L_{\rm env} \approx L_{\rm Edd}$ and $T_{\rm eff}\simeq 5000~{\rm K}$ as fiducial values, noting that the Eddington luminosity for $M_{\rm BH}\simeq10^{6.5}M_\odot$ is $L_{\rm Edd}\simeq 10^{44.6}~{\rm erg~s^{-1}} \sim 10^{11}L_\odot$, consistent with values considered in previous studies \citep[see][]{2025MNRASKido, Umeda_2026}. A low-density polar funnel is assumed to open from an inner radius $r_0$ to a covered radius $r_{\rm cov}$. The base of the funnel is defined as $r_0=r_{\min}R_g$, where $R_g=GM_{\rm BH}/c^2$ is the gravitational radius and $G$ is the gravitational constant. We set $r_{\min}=2$, consistent with the expectation that the funnel connects to the innermost accretion flow near the BH horizon. The covered radius $r_{\rm cov}$ represents the scale beyond which the envelope globally confines the jet. For simplicity, we assume that dissipation through internal shocks occurs in the vicinity of this region and take a fiducial dissipation radius of $r_{\rm dis}\approx r_{\rm cov}\simeq 10^{16}$ cm. The physical picture is analogous to choked jet models considered in the literature of jet-driven supernovae~\citep{Iocco:2007td,2013PhRvL.111l1102M} and tidal disruption events~\citep{2017ApJ...838....3S,Mukhopadhyay:2023mld}. 

To specify the funnel geometry, we need to model how the opening angle evolves with radius. Motivated by the CDAF dynamics, conservation of angular momentum leads to a gradual collimation of the funnel with radius \citep{2000ApJ...539..809Q, Meszaros+01coljet}:
\begin{equation}
\theta_j(r)=\theta_0\left(\frac{r}{r_0} \right)^{-1/2},
\label{eq:theta_scaling}
\end{equation}
where $\theta_0$ is the opening angle at $r=r_0$. Near the horizon, the effects of centrifugal force and pressure gradients in the CDAF are expected to be comparable, leading to $\theta_0\sim\mathcal{O}(1)$ \citep[e.g.,][]{2015MNRAS_Sadowski, 2018PASJ...70..108K}. We use $\theta_0=1.0$ as a fiducial value in this work, and the opening angle is collimated to $\theta_j (r_{\rm dis}) \simeq 9.7\times10^{-3}r_{\rm dis, 16}^{-1/2}M_{\rm BH, 6.5}^{1/2}(\theta_0/1.0) (r_{\rm min}/2.0)^{1/2}$.

As discussed later, in the fiducial case with $\theta_0\sim\mathcal O(1)$, the funnel remains confined and does not puncture the outer envelope. However, if the base opening angle is too narrow, the jet ram pressure would eventually exceed the envelope gas pressure at large radii, causing the jet to break out. Such a breakout would lead to X-ray and radio emission, which is not observed in LRDs \citep{Kocevski2023ApJ, Yue2024ApJ, Juodzbalis2024MNRAS, Ananna2024ApJ, Akins2025, Mazzolari2024arXiv, Maiolino2025MNRAS, Gloudemans2025ApJ, Perger2025A&A}. We evaluate the threshold base opening angle, $\theta_{0, \min}$, below which breakout occurs. A convenient approach is to apply the “no breakout” condition at a representative outer radius in the funnel. To quantify this “no breakout” condition, we approximate the jet ram pressure, envelope gas pressure, mass density, and sound speed as \citep[see e.g.,][]{1989ApJ...345L..21BMCD, 2003MNRAS.345..575M, BNP11a, MI13b}
\begin{align}
P_{\rm jet} &\approx \frac{L_j}{\pi (r_{\rm dis}\theta_j(r_{\rm dis}))^{2}c}, \label{eq:Pjet}\\
P_{\rm env} &\approx\rho(r_{\rm dis})C_{s}^2(r_{\rm dis}), \label{eq:Penv} \\
\rho(r)&\approx \frac{M_{\rm env}}{4\pi\ln\left({r_{\rm ph}}/{r_0}\right)}\frac{1}{r^3}, \label{eq:rho}\\
C_{s}^2(r)&\approx \frac{G M_{\rm BH}}{r}, \label{eq:Cs}
\end{align}
respectively, $M_{\rm env}$ is the envelope mass. We adopt $M_{\rm env}\approx M_{\rm BH}$ as a fiducial value in this work. These approximations assume a quasi-spherical mass distribution across logarithmic shells and a sound speed set by the local Keplerian potential, and these assumptions are satisfied by the envelope solution \citep[e.g.,][]{1997ApJ...489..573Loeb, 1998A&A...333..379Ulmer, 2025MNRASKido}. Requiring $P_{\rm jet}/ P_{\rm env} < 1$ and substituting $\theta_j(r) = \theta_0 (r/r_0)^{-1/2}$, we obtain 
\begin{equation}
\frac{4\ln(r_{\rm ph}/r_0)L_j r_{\rm dis}^3}{cr_0 GM_{\rm BH}M_{\rm env}\theta_0^{2}} \leq
1.
\end{equation}
This condition yields the minimum base opening angle
\begin{align}
\ \theta_{0,\rm min} &=
\left( \frac{4c\ln \left({r_{\rm ph}}/{r_0}\right) L_j r_{\rm dis}^{3}} {r_{\rm min} M_{\rm env} (G M_{\rm BH})^2} \right)^{1/2} \nonumber \\
&\simeq 0.47 ~ L_{j,44.6}^{1/2}r_{\rm dis, 16}^{3/2} M_{\rm BH,6.5}^{-1}M_{\rm env, 6.5}^{-1/2} \nonumber \\
&\times\left(\frac{r_{\rm min}}{2.0}\right)^{-1/2}\left(\frac{\ln(r_{\rm ph}/{r_{0}})}{10.3}\right)^{1/2},
\label{eq:theta0min}
\end{align}
which is lower than our fiducial choice $\theta_0\simeq 1.0$, thereby ensuring confinement by the envelope while allowing dissipation inside the funnel. Hydrostatic envelope solutions also allow smaller envelope masses \citep{2025MNRASKido}. If a smaller $M_{\rm env}$ is adopted, $\theta_{0,\min}$ increases, making jet confinement within the envelope more difficult.

The jet head propagation can be understood in the ram pressure balance between the jet and the surrounding envelope. Following Eqs.~(4), (5), and (6) of \cite{MI13b}, the dimensionless jet luminosity is estimated to be $\tilde{L} \approx L_j/(\pi (r_{\rm dis} \theta_j(r_{\rm dis}))^2 \rho(r_{\rm dis})c^3) \simeq 1.0\times10^{-5}~L_{j,44.6}r_{\rm dis, 16}^2 M_{\rm BH,6.5}^{-1}M_{\rm env, 6.5}^{-1}(\theta_0/1.0)^{-2}(r_{\rm min}/2.0)^{-1}$ $\left(\ln(r_{\rm ph}/{r_{0}})/{10.3}\right) \ll 1$, leading to a suppressed head velocity given by $\beta_h \approx \beta_j\tilde{L}^{1/2}\simeq 3.2\times10^{-3}~L_{j,44.6}^{1/2}r_{\rm dis, 16} M_{\rm BH,6.5}^{-1/2}M_{\rm env, 6.5}^{-1/2}(\theta_0/1.0)^{-1}(r_{\rm min}/2.0)^{-1/2}$ $\left(\ln(r_{\rm ph}/{r_{0}})/{10.3}\right)^{1/2}$, where $\beta_j\sim 1$ is the jet velocity normalized by $c$. From Eq.~(\ref{eq:Cs}), the sound speed at the dissipation region is $C_s(r_{\rm dis}) \simeq 2.0\times10^8~{\rm cm~s^{-1}}~r_{\rm dis,16}^{-1/2}M_{\rm BH,6.5}^{1/2}$, and the resulting Mach number is estimated to be $\mathcal{M} \approx c\beta_h/C_s \simeq 0.47~L_{j,44.6}^{1/2}r_{\rm dis, 16}^{3/2} M_{\rm BH,6.5}^{-1}M_{\rm env, 6.5}^{-1/2}(\theta_0/1.0)^{-1}(r_{\rm min}/2.0)^{-1/2}$ $\left(\ln(r_{\rm ph}/{r_{0}})/{10.3}\right)^{1/2}$, which is lower than unity. Thus, a strong forward shock is not expected to form, and we do not consider jet forward shocks in this work.

Since the funnel is filled with photons emitted from the surrounding accretion flow, shocks at the dissipation region may be radiation mediated. In radiation-mediated shocks, photons broaden the shock front and smooth out the velocity jump, which suppresses efficient cosmic-ray acceleration \citep{2008PhRvL.100m1101L, 2013PhRvL.111l1102M, 2020PhR...866....1L}. Then, we evaluate the comoving baryon density of the jet at the dissipation region to check whether the shock is radiation mediated. The comoving baryon density at the dissipation region is estimated to be
\begin{eqnarray}
n'_j &\approx& \frac{L_j}{2\pi (r_{\rm dis}\theta_j(r_{\rm dis}))^{2}\Gamma_j^{2} m_p c^{3}} \nonumber \\
&\simeq& 3.8\times10^7 ~{\rm cm^{-3}}~L_{j,44.6}r_{\rm dis,16}^{-1}M_{\rm BH,6.5}^{-1} \nonumber \\
&\times&\left(\frac{\Gamma_j}{2.0}\right)^{-2} \left(\frac{\theta_0}{1.0}\right)^{-2}\left(\frac{r_{\rm min}}{2.0}\right)^{-1},
\label{eq:nj_prime}
\end{eqnarray}
where $\Gamma_j$ is the jet Lorentz factor and $m_p$ is the proton rest mass. We adopt a mildly relativistic value of $\Gamma_j\simeq2$ in this work. Radiation pressure from the surrounding accretion flow regulates the jet acceleration. This can be illustrated using a CDAF scaling, where the temperature decreases with radius as $T \propto r^{-3/8}$,\footnote{This temperature scaling follows from standard CDAF assumptions: the convective energy flow satisfies $4\pi r^2 F_{\rm conv}=L_{\rm env}={\rm const}$, where $F_{\rm conv}$ is the convective energy flux, the flow is in hydrostatic equilibrium, and, because the pressure is dominated by radiation, $F_{\rm conv}\approx C_{s, \rm CDAF} P_{\rm rad}$, where $C_{s,\rm CDAF}\propto r^{-1/2}$ is the sound speed in the CDAF and $P_{\rm rad}$ denotes the radiation pressure.} implying a photon energy density $U_\gamma\propto T^4\propto r^{-3/2}$. A modest radial variation results in stronger radiation pressure at smaller radii and a gradual outward acceleration of the jet. When the jets become relativistic, the forward radiation pressure is boosted in the comoving frame, increasing radiation drag and limiting further acceleration, consistent with $\Gamma_j\sim{\rm a~few}$.

We examine the relevant length scales to clarify the structure of the dissipation region. We estimate the comoving magnetic field strength and the corresponding Larmor radius, and compare them with the funnel width and the radial length. Using the jet power and geometry at $r_{\rm dis}$, the comoving magnetic field strength is estimated to be
\begin{eqnarray}  
B'&=&\left(\frac{8\pi \epsilon_B L_j}{2\pi \big(r_{\rm dis}\theta_j(r_{\rm dis})\big)^2\Gamma_j^2\beta_jc} \right)^{1/2} \nonumber \\
&\simeq& 1.3\times10^{2}{~\rm G}~L_{j,44.6}^{1/2}r_{\rm dis,16}^{-1/2}
M_{\rm BH,6.5}^{-1/2} \epsilon_{B,-2}^{1/2}\nonumber \\
&\times&\left(\frac{r_{\min}}{2.0}\right)^{-1/2}
\left(\frac{\theta_0}{1.0}\right)^{-1}
\left(\frac{\Gamma_j}{2.0}\right)^{-1},
\end{eqnarray}
where $\epsilon_B$ is the fraction of jet power carried by the magnetic field. We take $\epsilon_B=0.01$ as a fiducial value, comparable to values used in conventional AGN jet studies \citep{2008MNRAS.385..283C, 2013ApJ...768...54B, 2024ApJ...967..104Z}. For a proton with comoving energy $\varepsilon'_p$, the Larmor radius is
\begin{equation}
r_L=\frac{\varepsilon_p'}{eB'} \simeq 2.6\times10^{13}~{\rm cm}~\left(\frac{\varepsilon_p'}{\rm EeV}\right)B^{\prime -1}_{2.1}.
\end{equation}
Since the funnel opening angle at the dissipation region is $\theta_j(r_{\rm dis})\simeq 9.7\times10^{-3}$, one finds
\begin{equation}
r_L < r_{\rm dis} \theta_j(r_{\rm dis}) \ll \frac{r_{\rm dis}}{\Gamma_j}.
\end{equation}
This ordering indicates that even for protons with $\varepsilon_p' \sim {\rm EeV}$, the Larmor radius remains below the funnel width, ensuring that particles can be accelerated efficiently. Moreover, $r_{\rm dis}/\Gamma_j \gg r_{\rm dis}\theta_j(r_{\rm dis})$, reflecting the strong collimation of the jet at this radius.

From Eq.~(\ref{eq:nj_prime}), we estimate the Thomson optical depth at the dissipation region as
\begin{align}    
\tau_{\rm jet} &\approx n'_j\sigma_{\rm T}\frac{r_{\rm dis}}{\Gamma_j} \nonumber \\
&\simeq 0.13~L_{j,44.6}M_{\rm BH,6.5}^{-1}\left(\frac{\Gamma_j}{2.0}\right)^{-3}
\left(\frac{\theta_0}{1.0}\right)^{-2}\left(\frac{r_{\rm min}}{2.0}\right)^{-1},
\label{eq:tauT_jet}
\end{align}
where $\sigma_{\rm T}$ is the Thomson cross section. Since $\tau_{\rm jet}< 1$, the shocks at the dissipation region are not radiation-mediated, meaning that the cosmic rays can be accelerated efficiently. We have also confirmed that, with the funnel width, $\tau_{\rm jet}\simeq n'_j\sigma_{\rm T} r_{\rm dis}\theta_j(r_{\rm dis}) \ll 1$.

With the jet environment characterized, we proceed to the neutrino production mechanism. Since the funnel interior is dilute in baryons, cosmic rays can mainly interact with photons supplied by the surrounding accretion flows. To examine whether cosmic rays undergo such interactions before escaping, we evaluate the photomeson optical depth, which is determined by the photon density and hence by the disk temperature at the dissipation region.

The disk temperature at the dissipation region is estimated by balancing the radiation-pressure gradient to gravity:
\begin{equation}
\frac13 aT^4(r) \approx \frac{GM_{\rm BH }\rho(r)}{r},
\label{eq:Prad}
\end{equation}
where $a$ is the radiation density constant. Then, we estimate the disk temperature at the dissipation region, $T_{\rm dis}\equiv T(r_{\rm dis}) $, as
\begin{align}
    T_{\rm dis} &\approx \left(  \frac{3GM_{\rm BH} \rho(r_{\rm dis})}{ar_{\rm dis}} \right)^{1/4} \nonumber \\
    &\simeq 1.7\times10^5 ~{\rm K}~r_{\rm dis,16}^{-1} M_{\rm BH,6.5}^{1/4}M_{\rm env,6.5}^{1/4} \nonumber \\
    &\times \left(\frac{\ln(r_{\rm ph}/{r_{0}})}{10}\right)^{-1/4}.
\end{align}
Using $T_{\rm dis}$, we evaluate the threshold proton energy for the photomeson producion in the jet comoving frame. Since the jet with Lorentz factor $\Gamma_j$ penetrates into this photon field, the threshold energy is given by  
\begin{align}
\varepsilon'_{{\rm thr},p\gamma} &\approx \frac{m_p m_{\pi^\pm} c^4}{2.8\Gamma_j k_B T_{\rm dis}} \nonumber\\
&\simeq1.6\times10^{15}~{\rm eV}~r_{\rm dis,16}~M_{\rm BH,6.5}^{-1/4}M_{\rm env,6.5}^{-1/4} \nonumber \\
&\times\left(\frac{\Gamma_j}{2.0}\right)^{-1}\left(\frac{\ln(r_{\rm ph}/r_{0})}{10.3}\right)^{1/4},
\label{eq:pg_threshold}
\end{align}
where $k_B$ is the Boltzmann constant and $m_{\pi^\pm}$ is the charged pion mass. This estimate implies that photomeson production with thermal-peak photons requires protons to reach $\sim{\rm PeV}$ energies in the jet comoving frame for the fiducial parameters.

We then evaluate the photomeson optical depth in the dissipation region. Due to the jet motion, the comoving photon number density is boosted to
\begin{align}
    n'_{\gamma } &\approx \Gamma_j \frac{ aT_{\rm dis}^4}{2.8k_BT_{\rm dis}} \nonumber \\ 
    &\simeq 1.9\times10^{17}{~\rm cm^{-3}}~r_{\rm dis, 16}^{-3}M_{\rm BH,6.5}^{3/4}M_{\rm env,6.5}^{3/4} \nonumber \\
    &\times \left(\frac{\Gamma_j}{2.0}\right)\left(\frac{\ln(r_{\rm ph}/{r_{0}})}{10.3}\right)^{-3/4}.
\end{align}
Given the photomeson cross section $\sigma_{p\gamma}\sim 5\times10^{-28}~{\rm cm^2}$, the optical depth for the photomeson production is
\begin{align}
    \tau_{p\gamma}&\approx n_\gamma'\sigma_{p\gamma}{r_{\rm dis}}{\theta_j(r_{\rm dis})} \nonumber \\
    &\simeq 9.0\times10^3~ r_{\rm dis,16}^{-5/2}M_{\rm BH,6.5}^{5/4}M_{\rm env,6.5}^{3/4} \nonumber \\
    &\times \left(\frac{\Gamma_j}{2.0}\right)\left(\frac{\theta_0}{1.0}\right)\left(\frac{r_{\rm min}}{2.0}\right)^{1/2}\left(\frac{\ln(r_{\rm ph}/{r_{0}})}{10.3}\right)^{-3/4}\gg 1.
\label{eq:ph_num_density}
\end{align}
Thus, once protons exceed the photomeson threshold in this copious thermal photon field, the photomeson production efficiency is almost $100\%$. 

In addition, a fraction of protons with energies $\varepsilon_p' \lesssim \varepsilon'_{{\rm thr},p\gamma}$ may escape the funnel and enter the dense envelope \citep{2013PhRvL.111l1102M}. Because of the high gas density, these protons can generate lower-energy neutrinos via inelastic $pp$ collisions $(p+p\rightarrow p+p+\pi)$ with nearly $100\%$ efficiency. However, the detection of these neutrinos is challenging since they lie in the sub-TeV energy range, where current neutrino detectors have limited sensitivity. Moreover, their contribution to the diffuse neutrino background is expected to be negligible, and thus we neglect $pp$-induced neutrinos in this work.

\section{Neutrino Emission from LRDs}\label{sec:neutrino}
Having established the funnel geometry in Section~\ref{sec:funnel}, we now turn to the neutrino emission expected from LRDs. In this section, we provide analytic estimates for the maximum proton energy and the associated neutrino luminosity (Section~\ref{sec:ana_estimate}), present the AMES calculation of the relevant timescales and the steady-state neutrino spectrum (Section~\ref{sec:AMES}), and examine neutrino escape and secondary pair effects on the dissipation region (Section~\ref{sec:neu_esc_&_sec_inj}).

\subsection{Analytic Estimates}\label{sec:ana_estimate}
First, we estimate the maximum energy of accelerated protons. The particle acceleration timescale at the jet comoving frame is phenomenologically estimated to be
\begin{eqnarray}
t_{\rm acc}' &\approx& \eta \frac{\varepsilon_p'}{eB'c} \nonumber \\
&\simeq & 8.7\times10^3 ~{\rm sec} ~\left(\frac{\varepsilon'_p}{\rm EeV}\right) \eta_1 L_{j,44.6}^{-1/2} r_{\rm dis,16}^{1/2}M_{\rm BH,6.5}^{1/2}\epsilon_{B,-2}^{-1/2} \nonumber \\
&\times&\left(\frac{r_{\min}}{2.0}\right)^{1/2}
\left(\frac{\theta_0}{1.0}\right)\left(\frac{\Gamma_j}{2.0}\right),
\label{eq:tacc}
\end{eqnarray}
where $\eta\sim10$ is the gyrofactor \citep[e.g.,][]{CS14c}. At the dissipation region, the photomeson optical depth is high ($\tau_{p\gamma}\gg1$), implying that $t'_{p\gamma}\ll r_{\rm dis}\theta_j(r_{\rm dis})/c\approx t'_{\rm esc}\approx t'_{\rm ad}$. Thus, both escape and adiabatic cooling are negligible, and photomeson production dominates the proton energy losses. As the photons supplied by the surrounding accretion flow are copious, we can neglect photons produced within the jet.\footnote{In the confined-funnel framework, $P_{\rm jet}<P_{\rm env}$ with $P_{\rm env}\approx aT_{\rm dis}^4/3$ (see Eqs.~(\ref{eq:Pjet}), (\ref{eq:Penv}), and (\ref{eq:Prad})), and the internally generated photon energy density cannot exceed $P_{\rm jet}$. Hence, photon energy density provided by the surrounding disk exceeds that produced within the jet, and we neglect the internal photon field in this work.} Then, the photomeson cooling timescale is roughly estimated to be $t_{p\gamma}'\approx 1/(n_\gamma' \sigma_{p\gamma}\kappa_{p\gamma}c)\simeq 1.8~{\rm sec} ~ r_{\rm dis, 16}^3 M_{\rm BH,6.5}^{-3/4}M_{\rm env, 6.5}^{-3/4}(\Gamma_{j}/2.0)^{-1}\left(\ln(r_{\rm ph}/{r_{0}})/{10.3}\right)^{3/4}$, where $\kappa_{p\gamma}\simeq0.2$ is the inelasticity for the photomeson production. 

Setting $t_{\rm acc}'=t_{p\gamma}'$, we obtain maximum proton energy as
\begin{align}
 \varepsilon'_{{p,\rm max},p\gamma}&\approx \frac{eB'}{\eta n'_\gamma \sigma_{p\gamma}\kappa_{p\gamma}} \nonumber \\
&\simeq2.1\times10^{14} {\rm~eV}~ 
\eta_{1}^{-1}L_{j,44.6}^{1/2}r_{\rm dis,16}^{5/2}M_{\rm BH,6.5}^{-5/4} \nonumber \\
&\times M_{\rm env,6.5}^{-3/4} \epsilon_{B,-2}^{1/2} \left(\frac{r_{\rm min}}{2.0}\right)^{-1/2}
\left(\frac{\theta_{0}}{1.0}\right)^{-1} 
\nonumber \\
&\times
\left(\frac{\Gamma_{j}}{2.0}\right)^{-2}
\left(\frac{\ln(r_{\rm ph}/{r_{0}})}{10.3}\right)^{3/4}.
\label{eq:Emax}
\end{align}
Since the analytic estimate gives $\varepsilon'_{p,\max}~<~\varepsilon'_{{\rm thr},p\gamma}$, the photomeson production cannot proceed with photons near the thermal spectral peak. Thus, the photomeson production mainly occurs with photons in the Wien tail, where the photon density is exponentially suppressed. The cooling time $t'_{p\gamma}$ becomes long, which shifts the photomeson--limited maximum proton energy to a higher energy.

Next, we estimate the neutrino luminosity produced through the photomeson production. Following the order-of-magnitude estimate introduced in Section~\ref{sec:diffuse_convert}, we assume a fraction $\epsilon_p \simeq0.1$ of the jet power to be carried by non-thermal protons with a $\varepsilon_p'^{-2}$ spectrum between $\varepsilon'_{p,\min}$ and $\varepsilon'_{p,\max}$:
\begin{equation}
\varepsilon_p'^2\frac{dN'_p}{d\varepsilon_p'dt'} \approx \frac{\epsilon_p  L'_j}{\ln\left(\varepsilon'_{p,\max}/\varepsilon'_{p, \min}\right)}
= \epsilon_p f_{\rm bol} L'_j,
\end{equation}
where $L'_j = L_j/\Gamma_j^2$ is the (beaming-corrected) jet luminosity in the comoving frame and $f_{\rm bol}^{-1} = \ln(\varepsilon'_{p,\max}/\varepsilon'_{p,\min})$ is the bolometric correction. For the fiducial parameters, we use $\varepsilon'_{p,\max}\simeq2.0\times10^{15}$ eV (see Fig.~\ref{fig:timescale} in the later subsection) and $\varepsilon'_{p,\min}\simeq m_pc^2$, and this evaluates to $\ln(\varepsilon'_{p,\max}/\varepsilon'_{p,\min}) \simeq 15$. Thus, the approximate choice $f_{\rm bol}\simeq0.1$ in Section~\ref{sec:diffuse_convert} is sufficiently justified. 

Since the photomeson production occurs in the Wien tail rather than at the thermal peak, we introduce an effective threshold energy $\varepsilon_{{\rm thr},p\gamma}^{' \rm eff} \equiv \varepsilon'_{{\rm thr},p\gamma}/2$, which reflects that the photomeson production begins with photons whose energies are effectively about twice the thermal peak. As $\tau_{p\gamma} \gg 1$, once protons exceed $\varepsilon^{' \rm eff}_{{\rm thr},p\gamma}$, the photomeson cooling timescale is the shortest among all relevant processes. Thus, in this regime, we do not introduce an explicit photomeson production efficiency factor $f_{p\gamma}(=t_{p\gamma}'^{-1}/t_{\rm cool}'^{-1})$, where $t_{\rm cool}'$ is the proton cooling timescale defined later. 

The bolometric all-flavor neutrino luminosity in the comoving frame is approximated to be
\begin{align}
L_\nu'^{\rm (bol)} &\approx \frac{3}{8} \epsilon_p  L'_j 
\frac{\ln\left(\varepsilon'_{p,\max}/\varepsilon^{' \rm eff}_{{\rm thr},p\gamma}\right)}{\ln\left(\varepsilon'_{p,\max}/\varepsilon'_{p,\min}\right)} \nonumber \\
&\simeq 2.4\times10^{41} ~ 
{\rm erg\ s^{-1}} ~L_{j,44.6}\epsilon_{p,-1} \left(\frac{\Gamma_j}{2.0}\right)^{-2}\nonumber \\
&\times\left(\frac{\ln\left(\varepsilon'_{p,\max}/\varepsilon^{' \rm eff}_{{\rm thr},p\gamma}\right)}{0.9}\right) \left(\frac{\ln\left(\varepsilon'_{p,\max}/\varepsilon'_{p,\min}\right)}{15}\right)^{-1}.
\label{eq:Lnubol}
\end{align}
In our fiducial setup, $\varepsilon'_{p,\max}/\varepsilon^{' \rm eff}_{{\rm thr},p\gamma}\sim{\rm a~few}$. If dissipation occurs at small radii, $\varepsilon'_{p,\max}$ takes a low value, the logarithmic factor is reduced, $L'_\nu$ drops, and the cumulative LRD contribution to the diffuse neutrino background will be less significant.

\subsection{Numerical Results with AMES}\label{sec:AMES}
\begin{figure}[tb]
    \centering
    \includegraphics[width=\linewidth]{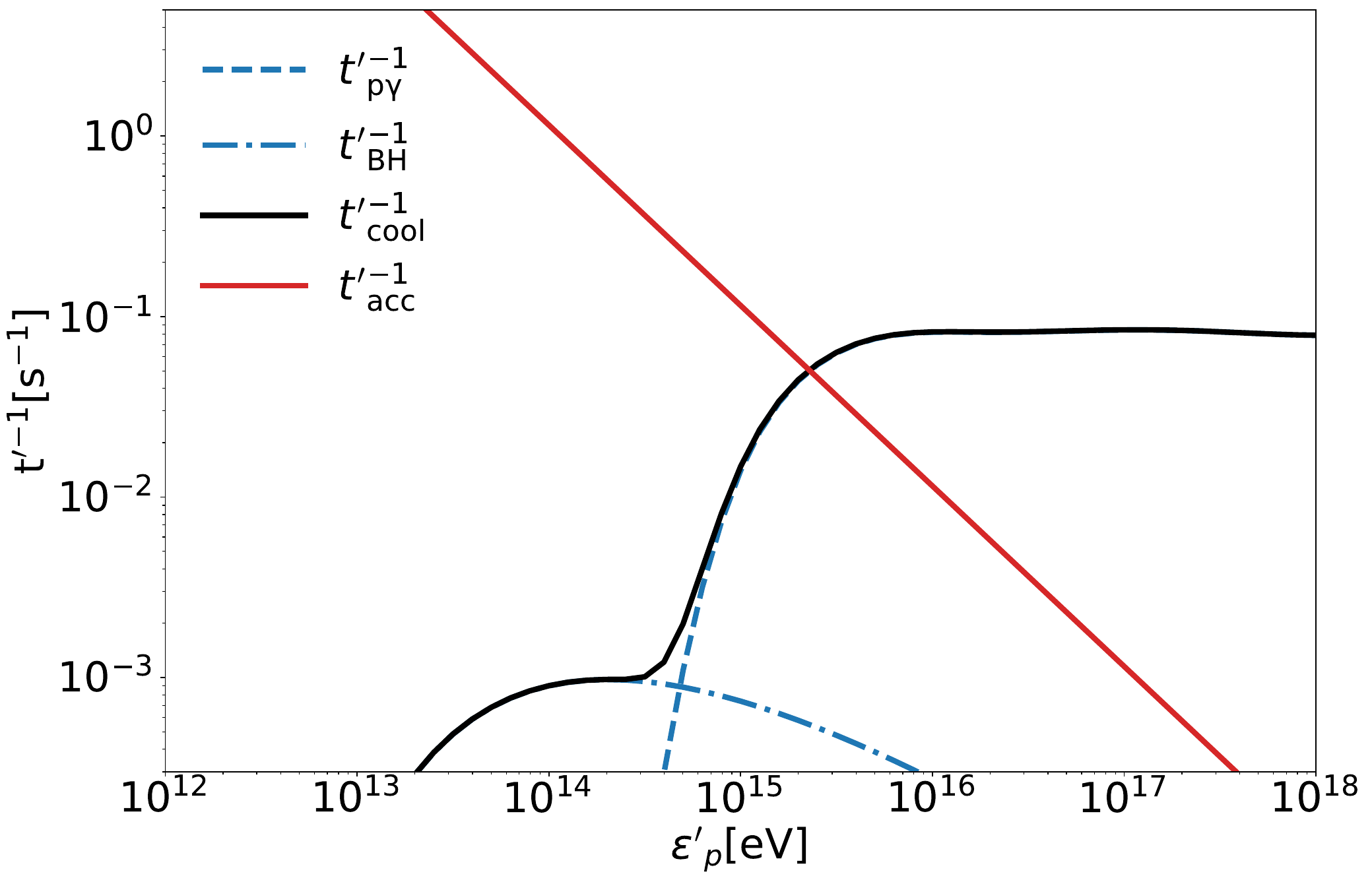}
    \caption{Acceleration and loss rates at the dissipation region in the comoving frame. Shown are $t_{\rm acc}'^{-1}$ (acceleration; red solid), $t_{p\gamma}'^{-1}$ (photomeson; blue dashed), $t_{\rm BH}'^{-1}$ (Bethe--Heitler; blue dash-dotted), and the total cooling rate $t_{\rm cool}'^{-1} = t_{ p\gamma}'^{-1}+t_{\rm BH}'^{-1}$ (black solid). The horizontal axis is the proton energy $\varepsilon'_p$, and the vertical axis is the inverse timescale $t'^{-1}$. All curves are computed with $r_{\rm dis}=10^{16}$~cm, $L_j=L_{\rm Edd}$ for a $10^{6.5}M_\odot$ BH, $\Gamma_j=2.0$, $\theta_0=1.0$, $\epsilon_p=0.1$, and $\epsilon_B=0.01$ (see Table~\ref{tab:params}).}
    \label{fig:timescale}
\end{figure}

\begin{figure}[tb]
    \centering
    \includegraphics[width=\linewidth]{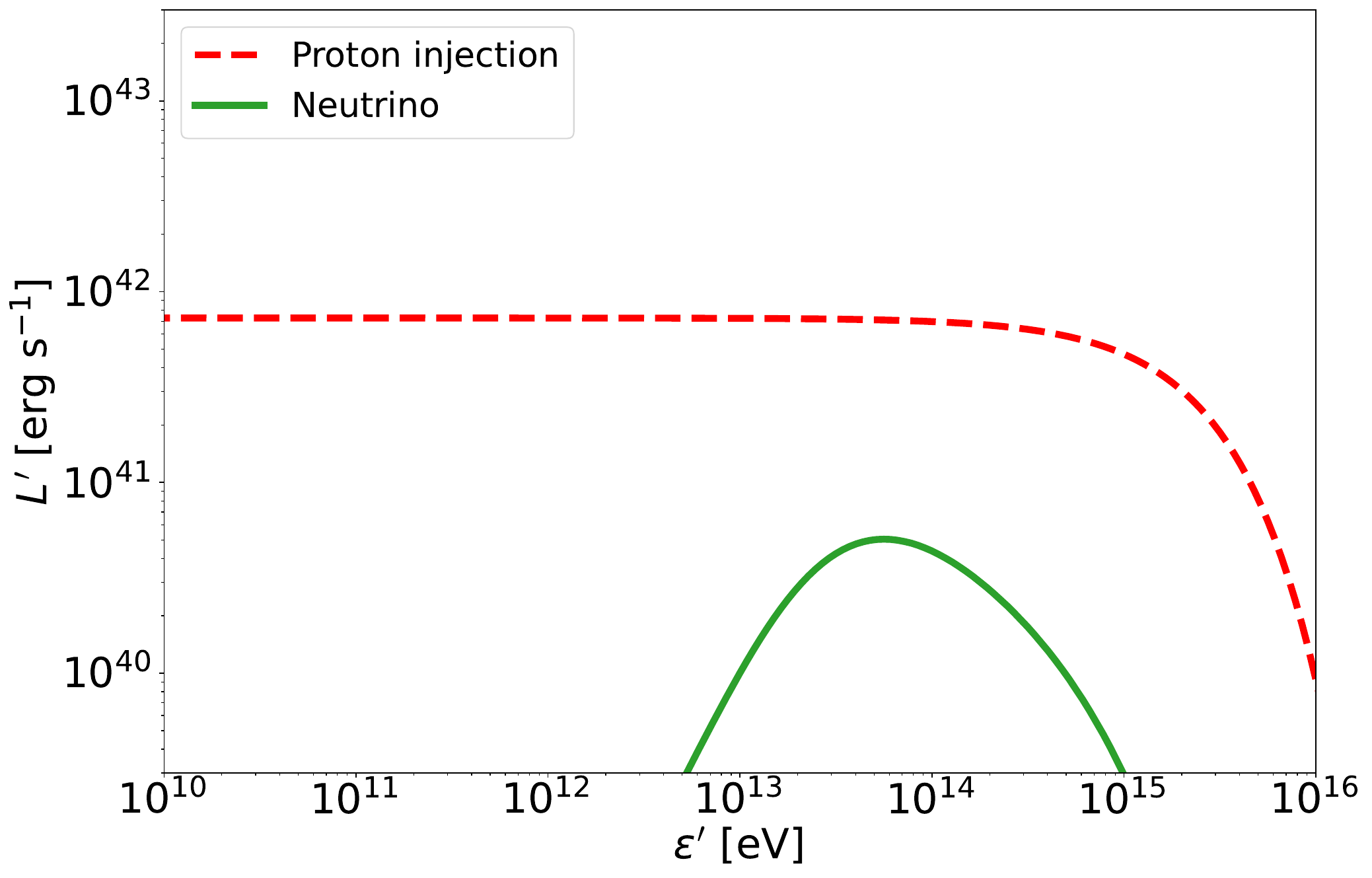}
    \caption{Comoving spectral luminosities $L'$ as functions of comoving energy $\varepsilon'$. The red dashed curve shows the proton injection spectrum, $L'_p(\varepsilon'_p)$, used for the neutrino calculation. The green solid curve shows the resulting all-flavor neutrino spectrum, $L'_\nu(\varepsilon'_\nu)$, computed with the AMES code using the fiducial parameters in Table~\ref{tab:params}, the same as in Fig.~\ref{fig:timescale}.}
    \label{fig:neu_Lum}
\end{figure}

\begin{table}[t]
\centering
\caption{Fiducial parameters adopted for AMES.}
\label{tab:params}
\begin{tabular}{rll}
\hline\hline
Parameter & Definition & Value \\
\hline
$M_{\rm BH}$        & Black hole mass                           & $10^{6.5}~M_\odot$ \\
$M_{\rm env}$       & Envelope mass                             & $10^{6.5}~M_\odot$ \\
$\epsilon_p$        & Proton acceleration efficiency            & $0.1$ \\ 
$\epsilon_B$        & Magnetic energy fraction                  & $0.01$ \\ 
$\Gamma_{j}$        & Jet Lorentz factor                        & $2.0$ \\
$\theta_{0}$        & Base opening angle                        & $1.0$ \\
$r_{\rm dis}$       & Dissipation radius                        & $10^{16}~ \rm{cm}$ \\
$r_{0}$             & Minimum radius                            & $2.0R_{g}$ \\
$\eta$              & Gyrofactor                                & $10$ \\
$\delta_D$          & Doppler factor  ($\theta_{\rm v} = 0$)    & $3.7$ \\
\hline
\end{tabular}
\end{table}

In this section, we present the results obtained by the AMES code. In the code, we solve the energy-coupled kinematic equations for protons, pions, muons, and electrons/photons with operators for continuous radiative losses (synchrotron/inverse Compton (IC)), escape, and injection, together with interaction terms for photomeson production, Bethe--Heitler process ($p+\gamma \rightarrow p+e^++e^-$), and inelastic $pp$ collisions \citep{2025arXiv251223231Wei}. The system is evolved until a steady state is reached, such that the secondary cooling, including muon/pion cooling and full electromagnetic cascades, is treated self-consistently. In this calculation, we use a fiducial parameter set for the BH-envelope-jet system, tabulated in Table~\ref{tab:params}. The adopted parameters are observationally motivated or consistent with previous studies and thus represent a conservative fiducial choice (see Section~\ref{sec:funnel}). The major uncertainty is the dissipation radius $r_{\rm dis}$, whose smaller values would reduce the neutrino contribution by lowering $\varepsilon'_{p,\max}$, as discussed in the previous subsection (see Eqs.~(\ref{eq:Emax}) and (\ref{eq:Lnubol})). 

For the target photon field, we consider the thermal radiation from the surrounding accretion flow. The comoving photon spectrum is \citep[see e.g.,][]{2019ApJ...887L..16K, 2023ApJ...950..190M}
\begin{equation}
\frac{{d}n'_{\gamma,\rm disk}}{{ d}\varepsilon'_\gamma} = \frac{2\pi}{h^3c^3}~\left(\frac{\varepsilon'_\gamma}{\Gamma_j}\right)^2\frac{1}{\exp\big({\varepsilon'_\gamma}/{(\Gamma_jk_BT_{\rm dis})}\big)-1},
\end{equation}
where $n'_{\gamma, \rm disk}$ is the comoving photon number density provided by the accretion flows, $\varepsilon'_{\gamma}$ is the comoving photon energy, and $h$ is the Planck constant. This spectrum is used in AMES to evaluate the $p\gamma$ interactions.

Fig.~\ref{fig:timescale} shows the relevant timescales for LRDs. We estimate the total cooling timescale as $t_{\rm cool}^{\prime -1} = t_{p\gamma}^{\prime -1} + t_{\rm BH}^{\prime -1},$ where $t'_{\rm BH}$ is the Bethe--Heitler cooling timescale. The acceleration timescale intersects the cooling timescale at $\varepsilon'_p \simeq 2.2\times 10^{15}$ eV, roughly consistent with our analytic estimate (Eq.~(\ref{eq:Emax})), especially when the exponential suppression in the Wien tail of the thermal photon spectrum is taken into account. At higher energies, photomeson production dominates the cooling, justifying our assumption of $f_{p\gamma}\simeq 1$ for $\varepsilon'_p \gtrsim \varepsilon^{' \rm eff}_{{\rm thr}, p\gamma}$. At lower energies, due to the abundant thermal photons, protons lose their energy through the Bethe–Heitler process. If the dissipation radius $r_{\rm dis}$ is small, the comoving photon number density is high, shortening both $t'_{\rm BH}$ and $t'_{p\gamma}$. In sufficiently compact regions, $t'_{\rm BH}$ could fall below $t'_{\rm acc}$ before photomeson cooling is dominant, leading to a Bethe–Heitler–limited maximum energy.

We show the all-flavor neutrino spectrum computed with AMES and the proton injection spectrum in Fig.~\ref{fig:neu_Lum}, plotted as a solid green line and a dashed red line, respectively. The resulting neutrino spectrum peaks at $L'_{\nu}\simeq 5.0\times10^{40}~\rm erg~s^{-1}$, in agreement with the analytic estimate within a factor of order unity (Eq.~(\ref{eq:Lnubol})). The spectrum peaks around $\varepsilon'_\nu \simeq 5.0\times10^{13}~\rm eV$, which corresponds to $\varepsilon'_p \simeq 1.0\times10^{15}~\rm eV$ with $\varepsilon'_\nu \approx 0.05\varepsilon'_p$. This indicates that the spectral peak occurs near the maximum proton energy, $\varepsilon'_{p,\max}$.

Finally, Doppler boosting rescales the spectrum. For an on-axis view, the Doppler factor is $\delta_D=1/(\Gamma_j(1-\beta_j \cos \theta_{\rm v}))$, where $\theta_{\rm v}\simeq 0.0 $ is the viewing angle. The emitted neutrino energy and beaming-corrected luminosity are boosted to $\varepsilon_\nu=\delta_D \varepsilon_\nu',~ L_{\nu}=\delta_D^{2} L_{\nu}'$. Taking into account the cosmological redshift of the neutrino energy, observed neutrino energy is given by $E_{\nu}= \varepsilon_\nu/(1+z)$, and thus, LRDs at $z\simeq4$--7 are expected to contribute to the diffuse neutrino background around $10^{4.5}~{\rm GeV}$, with $\varepsilon'_{p,\rm max}\simeq 2.2\times10^{15}$ eV and $\varepsilon'_\nu \approx 0.05 \varepsilon'_{p}$.

\subsection{Neutrino Escape and Pair Injection Effects}\label{sec:neu_esc_&_sec_inj}
One might expect that the high baryon density of the envelope makes it opaque to neutrinos, but, as shown below, the neutrino opacity is negligible. From Eq.~(\ref{eq:rho}), we estimate the column density of the envelope as $N_H\approx M_{\rm env}/(4\pi \ln(r_{\rm ph}/r_{0}) r_{\rm dis}^2 m_p)\simeq 2.9\times10^{29}~{\rm cm^{-2}}~r_{\rm dis, 16}^{-2}M_{\rm env, 6.5} (\ln (r_{\rm ph}/r_{0})/10.3)^{-1}$. For neutrino energies in the TeV–PeV range, the neutrino–nucleon cross section is $\sigma_{\nu N}\simeq10^{-35}$–$10^{-33}~{\rm cm^2}$ \citep{1998PhRvD..58i3009G, 2011PhRvD..83k3009C, 2024PhRvD.109k3001X}, which yields an optical depth of $\tau_{\nu N}\approx N_H\sigma_{\nu N}\simeq 10^{-5}$–$10^{-3}$. Thus, neutrinos can escape from the envelope without significant attenuation or scattering.

Photomeson production also produces gamma rays together with neutrinos. The resulting high-energy gamma rays are absorbed via the Breit–Wheeler process ($\gamma+\gamma\rightarrow e^++e^-$), with an enormous optical depth $\tau_{\gamma\gamma}\approx n_\gamma' \sigma_{\gamma\gamma} r_{\rm dis}\theta_j(r_{\rm dis})\simeq 2.4\times10^6r_{\rm dis,16}^{-5/2}M_{\rm BH,6.5}^{5/4}M_{\rm env,6.5}^{3/4}$ $\left({\Gamma_j}/{2.0}\right) \left({\theta_0}/{1.0}\right)\left({r_{\min}}/2.0\right)^{1/2}\left({\ln(r_{\rm ph}/{r_{0}})}/{10.3}\right)^{-3/4} \!\gg\!1$, where $\sigma_{\gamma\gamma}\approx 0.2\sigma_{\rm T}$ is the cross section for the Breit–Wheeler process. Such pair production, together with the Bethe–Heitler process, can inject additional $e^\pm$ pairs into the dissipation region. An analytic estimate (see Appendix~\ref{app:secondary_inj} for details) shows that the steady-state pair multiplicity increases the Thomson optical depth only at the level of a few percent relative to the original electrons. Therefore, even with secondary pair injection, the additional opacity is negligible, meaning that the shocks are not radiation mediated in our fiducial setup.

\section{Contribution to the diffuse neutrino background}\label{sec:diffuse_neu}
\begin{figure}[t]
    \centering
    \includegraphics[width=\linewidth]{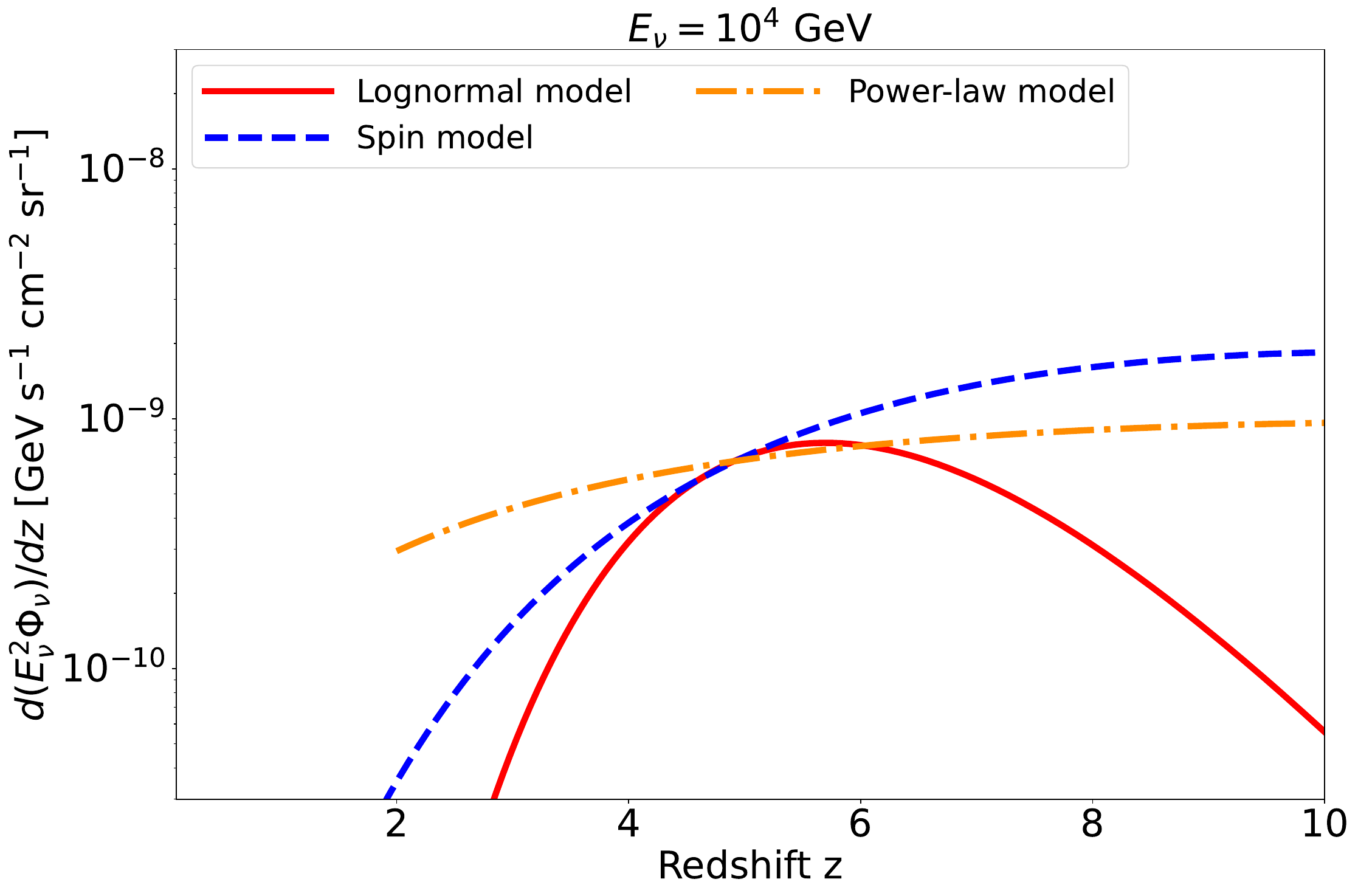}\\[0.5cm]
    \includegraphics[width=\linewidth]{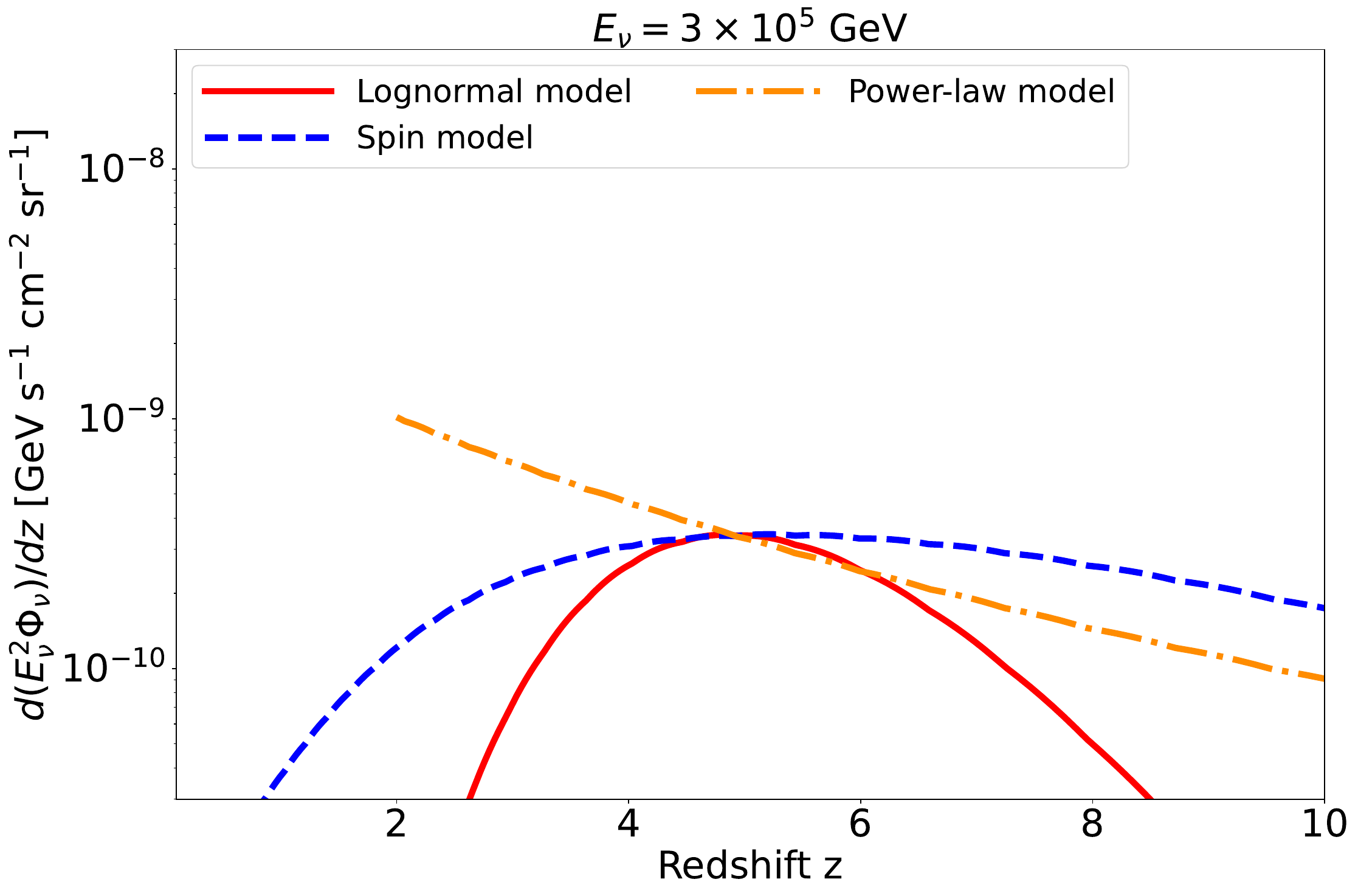}
    \caption{Redshift-differential diffuse neutrino intensity from LRDs, $d(E_{\nu}^2\Phi_{\nu})/dz$, for two observed energies. Top: $E_{\nu}=10^{4}~\rm GeV$. Bottom: $E_{\nu}=3\times10^{5}~\rm GeV$. Curves: red solid for the \textit{Lognormal model}, blue dashed for the \textit{Spin model}, and orange dot-dashed for the \textit{Power-law model}. We use the same parameters as in Fig.~\ref{fig:timescale} (see Table~\ref{tab:params}).}
    \label{fig:dFdz}
\end{figure}

\begin{figure}
    \centering
    \includegraphics[width=\linewidth]{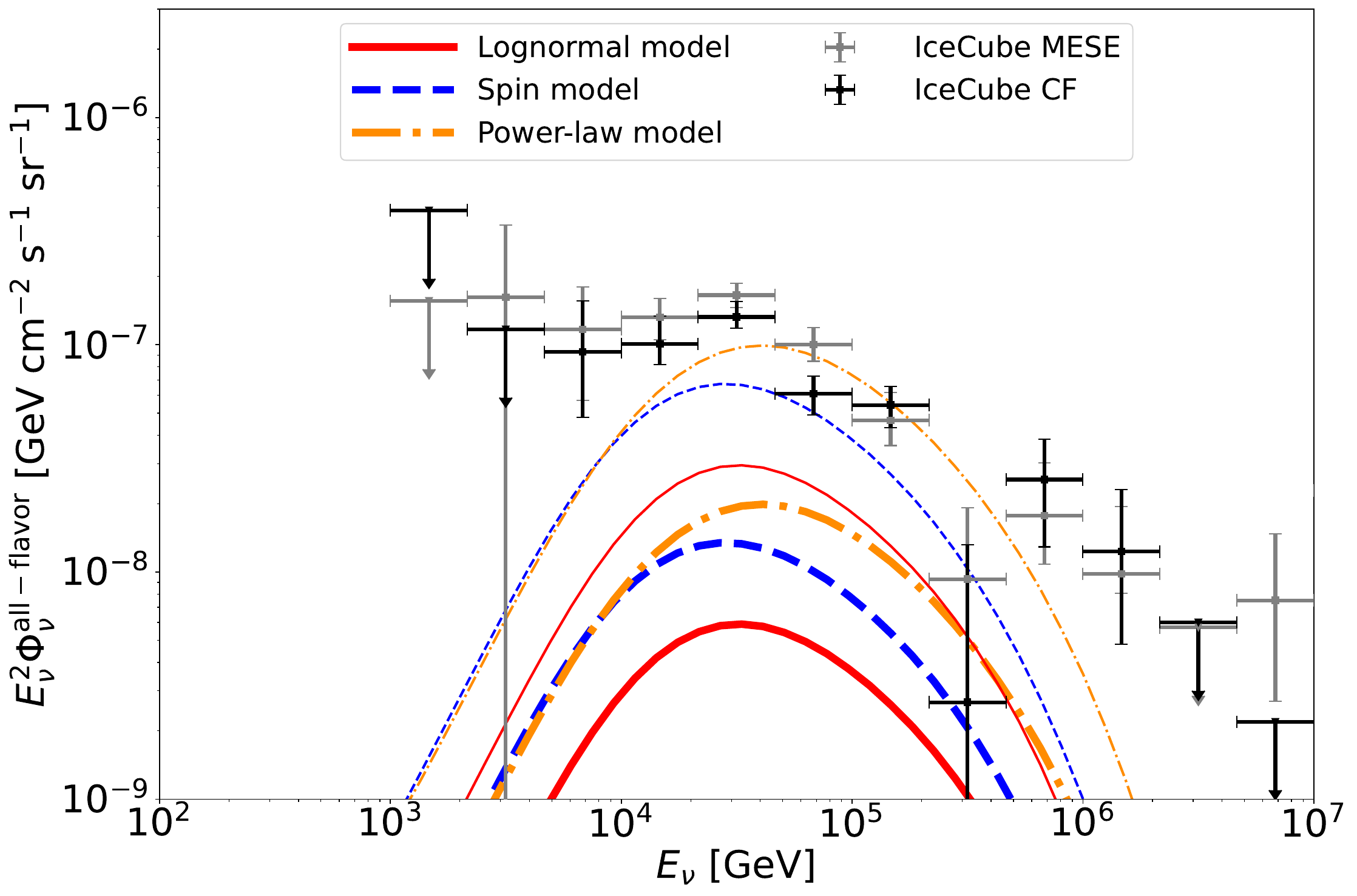}\\[0.5cm]
    \includegraphics[width=\linewidth]{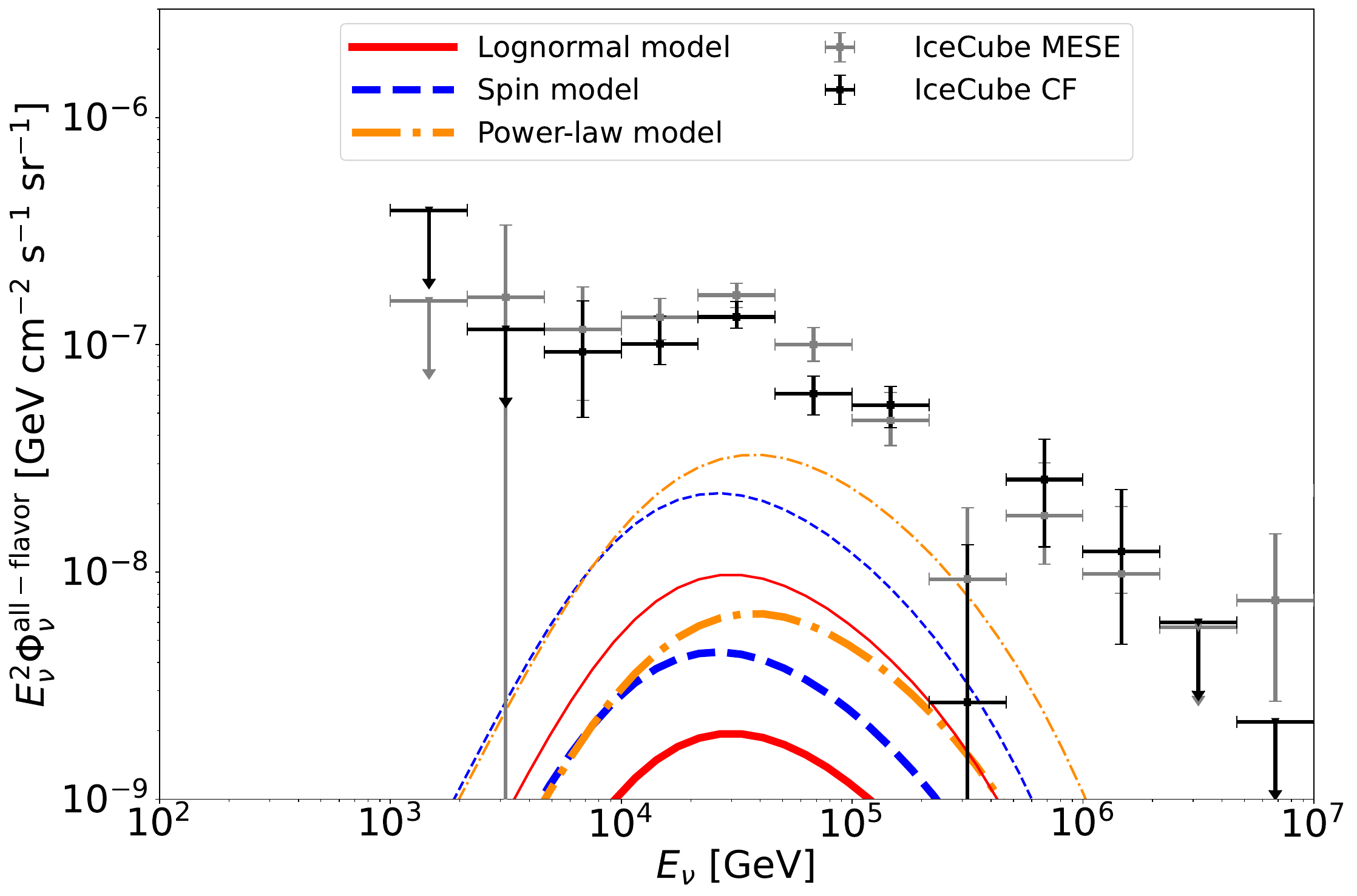}
    \caption{Diffuse all-flavor neutrino intensity expected from LRDs, compared with IceCube measurements. Top: $M_{\rm BH} = M_{\rm env}=10^{6.5}M_\odot,~ r_{\rm dis} = 10^{16}$ cm. Bottom: $M_{\rm BH} = M_{\rm env}=10^6M_\odot, ~r_{\rm dis} = 5\times10^{15}$ cm. All other parameters follow Table~\ref{tab:params}, and the line styles and colors are the same as in Fig.~\ref{fig:dFdz}. Thin lines with the same styles indicate the predictions obtained by adopting the same population models with $\epsilon_p = 0.5$. Gray and black points with error bars are taken from \citet{IceCube2025A}, showing the IceCube results from the MESE and the combined-fit (CF) analyses, respectively.}
    \label{fig:diffuse-Neu}
\end{figure}

In this section, we discuss the potential contribution of LRDs to the diffuse neutrino background. We specifically focus on a fiducial class of LRDs hosting BHs with mass $M_{\rm BH}\simeq10^{6.5}M_\odot$ and envelope mass $M_{\rm env}\simeq10^{6.5}M_\odot$, and assume that such sources are isotropically distributed across the Universe.

As the comoving number density of LRDs can be inferred from both observations and theoretical predictions, we adopt three representative models to capture these possibilities (see Appendix~\ref{app:numberdensity} for details): (i) the \textit{Lognormal model}, derived from the observed LRD population \citep{2025ApJ...988L..22I}, which provides the most conservative estimate; (ii) the \textit{Spin model}, an analytic estimate based on the probability distribution function of the halo spin parameter \citep{2025ApJ...989L..19P}, representing a theoretically allowed population; and (iii) the \textit{Power-law model}, normalized to the \textit{Spin model} at $z=5$, with $n(z)\propto(1+z)^{3}$, serving as a phenomenological benchmark for comparison. For each case, we compute the neutrino flux at each redshift bin and sum their contributions. 

We confirm that the adopted LRD population models remain consistent with the bolometric BH mass density recently inferred by \citet{Umeda_2026} once the current systematic uncertainties are taken into account, and these models also satisfy the constraints from the observed cosmic infrared background. These conclusions hold even under the optimistic assumption of $M_{\rm BH}=10^{6.5}M_\odot$. An order-of-magnitude comparison is presented in Appendix~\ref{app:numberdensity}.

We calculate the all-flavor diffuse neutrino intensity by integrating the differential redshift contribution. The diffuse neutrino intensity from LRDs at a given redshift $z$ is estimated to be
\begin{equation}
\frac{d}{dz}\left[{E^2_{\nu} \Phi_{\nu}}\right]
 =\frac{L_\nu\big((1+z)E_{\nu}\big)}{4\pi d_L^2(z)} n(z)\frac{dV_c}{dzd\Omega}.
\label{eq:dIdz}
\end{equation}
The all-flavor diffuse neutrino intensity is then obtained by integrating over redshift:
\begin{eqnarray}
E_{\nu}^2\Phi_{\nu}= \int_{z_{\min}}^{z_{\max}}\frac{d}{dz}\left[{E^2_{\nu}} \Phi_{\nu}\right] dz.
\label{eq:diffuse}
\end{eqnarray}
In this work, we take $z_{\min}=0.1$ and $z_{\max}=10$. This wide range is motivated by recent reports of LRD candidates in the nearby Universe \citep{2025arXiv250710659L} as well as possible detections at $z\simeq10$ \citep{2025arXiv250800057T}. Although the actual population across the entire range is uncertain, this assumption tests the potential contribution if such sources exist both locally and at high redshifts. For \textit{Power-law model}, extending the population unchanged to low redshift will overproduce the local number density and conflict with the scarcity of reported local detections. To remain conservative, we impose a low-$z$ cutoff for the \textit{Power-law model} and adopt $z_{\min,{\rm cut}}\simeq 2.0$, while other models are evaluated over the full $z_{\min}$–$z_{\max}$ range. 

Fig.~\ref{fig:dFdz} presents the redshift distribution of the diffuse neutrino intensity with the top panel showing $E_{\nu}=10^{4}$~{\rm GeV} and the bottom panel showing $E_{\nu}=3\times10^{5}$~{\rm GeV}. When the observed energy is below the intrinsic spectral peak, redshifting shifts emission from high redshift toward the peak, enhancing the contribution from distant LRDs despite the large luminosity distance. Thus, the intensity at a given redshift is higher than in the higher–energy case. Above the peak, the dominant contribution moves to low redshift, and the result depends on how slowly $n(z)$ declines toward low redshift. In the \textit{Lognormal model}, $n(z)$ peaks at $z\simeq5$--7, and the intensity concentrates near this peak in both panels. The \textit{Spin model} contains many LRDs at high redshift but very few in the nearby Universe, resulting in a nearly flat profile for $z\gtrsim4$ and a suppressed intensity at low redshift. The \textit{Power-law model} provides a high-redshift contribution similar to that of the \textit{Spin model}, and $n(z)$ remains relatively high toward low redshift, resulting in high intensity from nearby LRDs. This trend is apparent at higher observed energies.

These differences are reflected in the integrated diffuse neutrino intensities shown in Fig.~\ref{fig:diffuse-Neu}. The \textit{Lognormal model}, whose population is concentrated around $z\simeq5$–7 and suppressed elsewhere, reaches only $\sim10\%$ across $E_{\nu}\simeq 1.0\times10^{4}\text{--}5.0\times10^{4}~\rm GeV$, where redshifting brings the intrinsic neutrino spectral peak into this band, and the contribution is negligible at other energies. For the \textit{Spin model}, which retains more LRDs than the \textit{Lognormal model} at high redshifts, the contribution reaches $\sim20\%$ around $1.0\times10^{4}~{\rm GeV} \lesssim E_{\nu} \lesssim 1.0\times10^{5}~\rm GeV$. The \textit{Power-law model}, which has more LRDs at low redshifts, can contribute up to $\sim 30\%$ of the IceCube diffuse measurement over the same energy range and continues to contribute at the same level near $E_{\nu}\simeq 2.0\times10^5$ GeV. These estimates adopt $M_{\rm BH} \simeq 10^{6.5} M_\odot$ for all LRDs and assume that each LRD launches jets, and hence the thick curves represent optimistic upper limits. 

Thin lines in Fig.~\ref{fig:diffuse-Neu} show the corresponding predictions when the proton acceleration efficiency is increased to $\epsilon_p = 0.5$. This “pushed” case is included to demonstrate how strongly the proton acceleration efficiency needs to be enhanced to reproduce the observed diffuse neutrino background. The comparison with the IceCube data indicates that matching the diffuse intensity requires an efficiency of order $\epsilon_p\simeq0.5$. Nevertheless, both thick and thin curves imply that LRDs will be a non-negligible contributor to the diffuse neutrino background.

Recent infrared observations suggest that LRDs may lack significant dust \citep{Setton2025arXiv, Xiao2025A&A, 2025arXiv250905434G}, although other studies argue for the presence of dust \citep{Barro2024arXiv, Li2025ApJ, delvecchio2025arXiv}, and the current situation remains uncertain. If dust is absent, the bolometric luminosities inferred under the assumption of dust obscuration would be overestimated, and the BH mass could be as low as $M_{\rm BH}\simeq10^{6}M_\odot$. We calculate the contribution to the diffuse neutrino background with $M_{\rm BH}\simeq 10^{6}M_\odot$ and show the results in the bottom panel of Fig.~\ref{fig:diffuse-Neu}. For the low-mass case with $M_{\rm BH}=M_{\rm env}=10^{6}M_\odot$, we use a smaller dissipation radius of $r_{\rm dis}=5\times10^{15}~{\rm cm}$ to avoid jet breakout from the envelope (see Eq.~(\ref{eq:theta0min})), while keeping all other parameters fixed (see Table~\ref{tab:params}). In this setup, the jet luminosity and envelope temperature are both reduced, but the decrease in $r_{\rm dis}$ mitigates these effects, leading to only a modest change in the comoving thermal--photon number density. Consequently, $\varepsilon'_{p,{\rm max}}$ decreases by at most a factor of a few, and the predicted neutrino spectrum shifts to slightly lower energies with a reduced luminosity. As a result, LRDs with $M_{\rm BH}\simeq10^{6}M_\odot$ contribute at most $\sim10\%$ in the same energy range as the $M_{\rm BH}=10^{6.5}M_\odot$ case. Under the conservative \textit{Lognormal model}, their contribution to the diffuse background remains minor.

\section{Discussion}\label{sec:discussion}
\subsection{Detectability for individual LRDs}\label{sec:detectability}
Detecting individual LRDs through neutrino observations is challenging even with the improved sensitivity of upcoming neutrino detectors. Typically, an LRD has $L'_\nu\simeq5.0\times10^{40}~{\rm erg~s^{-1}}$ and $\delta_D\simeq3.7$ (see Fig.~\ref{fig:neu_Lum} and Table~\ref{tab:params}). LRDs are abundant at $z\simeq6$, which corresponds to a luminosity distance of $d_L\simeq58~{\rm Gpc}$. This gives a number flux at $E_{\nu}\simeq3.0\times10^4~{\rm GeV}$ of $\phi_{\nu}\approx \delta_D^4 L'_\nu/ (4 \pi d_L^2 E_{\nu}) \simeq3.9\times10^{-19}~{\rm cm^{-2}~s^{-1}}~L'_{\nu,40.7} (\delta_D/3.7)^4 d_{L,29.3}^{-2}$. The expected event rate is $ R_1 \approx \phi_{\nu} A_{\rm eff} \simeq 2.0\times10^{-16}{\text{--}}3.9\times10^{-16}~{\rm s^{-1}}~L'_{\nu,40.7} (\delta_D/3.7)^4 d_{L,29.3}^{-2}$, where $A_{\rm eff}\simeq5\times10^2\text{--}10^3~{\rm cm^2}$ is the effective area for IceCube--Gen2 at $E_{\nu}\simeq3.0\times10^4~{\rm GeV}$, a range obtained by scaling the effective area for IceCube by a factor of five at the same energy \citep{2021PhRvD.104b2002A, Gen2_2021}. This corresponds to $\simeq8.1\times10^{7}\text{--}1.6\times10^{8}~\rm yr$ per event, and thus a single LRD is undetectable as a point source.

Stacking analyses also do not help due to the large number of LRDs within the error region. A simple estimate illustrates this source confusion. Considering a typical error radius of $\theta_{\rm err}\sim1~{\rm deg}$ at $z\simeq 6$, the comoving volume of an error region is $V_{\rm err}\approx \pi(\theta_{\rm err} d_c(6))^2\Delta d_c \simeq 2.7\times10^{7}~{\rm cMpc^3}$, where $\Delta d_c\approx d_c(6.5)-d_c(5.5)\simeq4.2\times10^{2}~{\rm cMpc}$ is the corresponding comoving depth. With the LRD comoving number density of $n\simeq5\times10^{-5}~{\rm cMpc^{-3}}$ at $z\simeq 6$, the number of LRDs inside a single error region is $N_{\rm LRD}\approx nV_{\rm err} \simeq 1.4\times10^{3}$. As a result, each IceCube or IceCube--Gen2 error region contains many LRDs. This source confusion prevents a meaningful stacking strategy, because the expected neutrino flux per source is small and the candidate set within each error region is degenerate.

\subsection{Observational tests to distinguish LRDs from radio-quiet AGNs}\label{sec:neu-flavor}
Among the existing source classes, Seyfert-like, radio-quiet AGNs are the most relevant point of comparison in terms of energy range, since radio-quiet AGNs are also expected to contribute to the diffuse neutrino background in the $10^{4}$--$10^{5}$ GeV band \citep{Murase:2019vdl}. This similarity motivates an investigation of possible differences, which we discuss in this subsection.

Multiplet detection could distinguish the two scenarios. In Fig.~\ref{fig:rho-Lnu}, $\xi_z n_0^{\rm eff}$ of both classes lies close to the diffuse--requirement curve. On the other hand, the LRD point lies below the thick multiplet--limit curves, whereas the radio-quiet AGN point exceeds the thin SFR-based multiplet--limit curves for IceCube--Gen2. The redshift evolution factor $\xi_z$ for radio-quiet AGNs closely matches that for the SFR, which leads the corresponding multiplet--limit curves to approximately coincide with the SFR-based curves shown in Fig.~\ref{fig:rho-Lnu}. Therefore, if IceCube--Gen2 detects neutrino multiplets associated with radio-quiet AGNs, such observations would suggest that radio-quiet AGNs contribute substantially to the diffuse neutrino background in the $10^{4}$--$10^{5}$~GeV range, whereas LRDs do not.

Another distinctive feature of the LRD scenario is the suppression of muon-decay neutrinos, as muons are efficiently cooled by IC cooling in the copious photon field. From a muon cooling timescale calculation, we find that muons cool efficiently above $\varepsilon'_\mu \simeq 1.0\times10^{15}$ eV, corresponding to $\varepsilon'_\nu \simeq 3.3\times10^{14}$ eV. This suppression is already reflected in the neutrino luminosity shown in Fig.~\ref{fig:neu_Lum}. However, pion-decay neutrinos remain, and the overall reduction is limited to a factor of a few. \footnote{IC cooling also works for pions, but in the present case, pion IC cooling is suppressed by relativistic kinematic effects, whereas muon IC cooling remains mostly in the Thomson regime. As a result, the ratio of the cooling energies between pions and muons is larger than in the synchrotron cooling case. Since muon IC cooling already occurs near $\varepsilon'_{p,\max}$, neutrino suppression due to pion IC cooling is not significant in the neutrino spectrum.}

For $\varepsilon'_\nu \lesssim 3.3\times10^{14}$ eV, muons decay before cooling, and the resulting flavor composition at the source is the conventional pion-decay ratio, $(\nu_e:\nu_\mu:\nu_\tau) \simeq (1:2:0)$, which after oscillations approaches $(1:1:1)$ at Earth. In contrast, for $\varepsilon'_\nu \gtrsim 3.3\times10^{14}$ eV, muon decays are suppressed, leaving only $\nu_\mu$ from direct pion decay. Then, the flavor ratio is $(0:1:0)$, which oscillates to approximately $(1:1.8:1.8)$ at Earth \citep[e.g.,][]{2005PhRvL..95r1101K, 2015PhRvD..92h5047K}.
  
This energy-dependent transition in flavor composition provides a distinctive observational signature of the LRD scenario. In particular, the diffuse neutrino background in $E_{\nu} \simeq 1.0\times10^4 \text{--} 2.0\times10^{5}~\rm GeV$ could receive a $\sim 30\%$ contribution from LRD. Future measurements with IceCube-Gen2 \citep{Aartsen:2019Gen2}, KM3NeT \citep{KM3NeT16a}, Baikal-GVD \citep{2014NIMPA.742...82A}, TRIDENT \citep{2022arXiv220704519Y}, P-ONE \citep{2020NatAs...4..913A}, and HUNT \citep{Huang:2023mzt} will test this prediction and may determine whether the diffuse neutrino background is primarily powered by radio-quiet AGN or by LRDs by identifying this energy-dependent feature \citep[see e.g.,][]{2023arXiv230815220L}.

\subsection{LRD environment compared with AGN jets}\label{sec:AGN-comparison}
The physical setup considered in this work is close to AGN jet systems, but the role of the surrounding medium makes the LRD scenario fundamentally different. In AGN jets, the jet largely determines its own dynamics, including the opening angle and propagation to large distances \citep[e.g.,][]{Mck06a, TNM11a, 2019ApJS..243...26P}. Internal photons generated within the jet provide the main targets for cosmic-ray interactions, and both gamma rays and neutrinos can escape from the system \citep[e.g.,][]{mid14, 2023ecnp.book..483M}. 

In LRD jets, by contrast, the funnel opening angle and the jet dynamics are externally imposed by the surrounding disk. The surrounding accretion disk supplies a dense photon field, which prevents free escape of cosmic rays and leads to energy losses dominated by photomeson production. The resulting internal photons are absorbed through the Breit--Wheeler process, and the optically thick envelope further suppresses photon escape, leaving neutrinos as the only viable messengers. These features distinguish LRD jets from ordinary AGN jets in both dynamics and messenger channels.

The surrounding envelope and disk determine the LRD jet dynamics, yet key parameters remain uncertain, including the envelope mass, the base opening angle of the funnel, and the jet Lorentz factor set by acceleration due to the radiation pressure from the accretion flow. Addressing these issues requires radiation-magnetohydrodynamic simulations of jet propagation through a CDAF background. These simulations are beyond the scope of this work, but will be essential for a complete understanding.

\section{Conclusion}\label{sec:conclusion}
In this work, we have proposed LRDs, a population of compact red galaxies recently revealed by JWST at high redshift, as a promising new class of high-energy neutrino emitters. A BH-envelope model, in which a SMBH is embedded in a dense medium, has been suggested for these objects. However, the envelope alone is collisional and cannot efficiently accelerate cosmic rays. Finite angular momentum in the inflow naturally carves a low-density polar funnel. We consider that a jet launched by the central BH propagates and dissipates in this funnel, where a copious photon field exists. Cosmic rays are efficiently cooled by the photomeson production, which produces both gamma rays and neutrinos. The resulting gamma rays are absorbed or scattered by the dense photon field and the surrounding envelope, leaving neutrinos as the dominant high-energy messenger from LRDs. This makes LRDs natural ``hidden neutrino sources.’’ Motivated by this picture, we evaluated their cumulative contribution to the diffuse neutrino background. 

Interestingly, our setup is a modern version of the BH-cocoon model proposed by \citet{Berezinsky1977} but we stress that these physical setups are entirely different from the AGN disk-corona model, in which non-thermal X-rays escape \citep[e.g.,][]{Murase:2019vdl,Kimura:2020thg}. We also note that choked jets embedded in the BH envelope with a similar density structure are considered in the context of tidal disruption events~\citep{2017ApJ...838....3S,Mukhopadhyay:2023mld}, and the bare BH model in \citet{Berezinsky1977} resembles such a stellar disruption rather than the AGN disk.

We calculated the neutrino emission from LRDs with the AMES code and evaluated their contribution to the diffuse neutrino background using three representative population models: a \textit{Lognormal model}, a \textit{Spin model}, and a simple \textit{Power-law model}, with $z_{\min}=0.1$ and $z_{\max}=10$. For the \textit{Power-law model}, we imposed a conservative low-redshift cutoff $z_{\min,{\rm cut}}=2.0$. Because LRDs are abundant at high redshift, their contribution to the diffuse neutrino background is dominated by sources at those epochs, as shown in Fig.~\ref{fig:dFdz}. We observe them after redshifting, and the resulting diffuse neutrino intensity is enhanced below the neutrino spectral peak (see the top panel of Fig.~\ref{fig:dFdz}).

In Fig.~\ref{fig:diffuse-Neu}, the \textit{Lognormal model}, whose population concentrates at $z\simeq5$--7, reaches at most $\sim10\%$ over $1.0\times10^{4}\text{--}5.0\times10^{4}~{\rm GeV}$ and remains subdominant elsewhere. By contrast, the \textit{Spin} and \textit{Power-law} models retain higher comoving densities at high redshift and yield a non-negligible fraction of the IceCube flux across $1.0\times10^{4}~{\rm GeV} \lesssim E_{\nu} \lesssim2.0\times10^{5}~{\rm GeV}$. We also examined the case with $M_{\rm BH}=M_{\rm env}=10^{6}~M_\odot$, in which only the \textit{Spin} and \textit{Power-law} models contribute at the $\sim 10\%$ level over the same energy range.

In the same energy range, radio-quiet AGNs are also expected to contribute to the observed diffuse neutrino background \citep{Murase:2019vdl}. Because LRDs lie below the multiplet threshold curves, multiplet searches are not feasible for them, whereas radio-quiet AGNs can exceed these curves (see Fig.~\ref{fig:rho-Lnu}). Thus, if multiplet events are observed from radio-quiet AGNs, such a detection can distinguish the two scenarios and indicate that the radio-quiet AGNs are the dominant contributors to the diffuse neutrino background. 

Another distinctive feature is the flavor composition of the diffuse neutrino background. The key point is the fate of secondary muons. The IC cooling is efficient above $\varepsilon'_\mu \simeq 1.0\times10^{15} ~ \rm{eV}$, corresponding to a neutrino energy of $\varepsilon'_\nu \simeq 3.3\times10^{14} ~ \rm{eV}$. As a result, muon-decay neutrinos are strongly suppressed at high energies, and the flavor ratio observed at Earth transitions from $(\nu_e:\nu_\mu:\nu_\tau) \simeq (1:1:1)$ at lower energies to $(1:1.8:1.8)$ at higher energies. Detecting such an energy-dependent shift would provide a distinctive and testable signature of the LRD scenario. Future measurements of the diffuse neutrino flavor composition at these energies could serve as a critical test of the scenario.

In our framework, photons generated in the jet are efficiently absorbed and scattered within the envelope, leaving neutrinos as the only detectable high-energy messengers. This feature provides a clear observational test. If alternative scenarios that predict electromagnetic counterparts are disfavored by continued non-detections in future radio and X-ray observations, the envelope-regulated LRD picture will be correspondingly strengthened.

\begin{acknowledgments}
We thank Masaru Shibata, Kazumi Kashiyama, and Jonathan Granot for their fruitful comments. We also thank Bing Theodore Zhang for guidance on the use of the AMES code. This work was developed based on discussions during the YITP workshop ``Exploring Extreme Transients: Frontiers in the Early Universe and Time-Domain Astronomy'' (YITP-W-25-08) at the Yukawa Institute for Theoretical Physics, Kyoto University. We are grateful to the organizers and participants for stimulating discussions. The authors also thank the participants of the Yukawa International Seminar YKIS2026a on ``Black Holes and Neutron Stars with Multi-Messengers''. Discussions during this long-term workshop were helpful in finalizing this work. R.K. is a Yukawa Fellow supported by Yukawa Memorial Foundation. This work is supported by JSPS KAKENHI Nos. 23H01172, 23H05430, 23H04900, 22H00130 (K. Ioka).
The work of K.M. was supported by the NSF Grants No.~AST-2108466, No.~AST-2108467, and No.~2308021.
S.S.K. acknowledges support by KAKENHI Nos. 22K14028, 21H04487, 23H04899, and the Tohoku Initiative for Fostering Global Researchers for Interdisciplinary Sciences (TI-FRIS) of MEXT’s Strategic Professional Development Program for Young Researchers.
K. Inayoshi acknowledges support from the National Natural Science Foundation of China (12573015, W2532003, 1251101148, 12233001), the Beijing Natural Science Foundation (IS25003), and the China Manned Space Program (CMS-CSST-2025-A09).
\end{acknowledgments}

\appendix
\section{LRD Population Models and Energetic Consistency}\label{app:numberdensity}
\begin{figure*}
    \centering
    \includegraphics[width=1.05\linewidth]{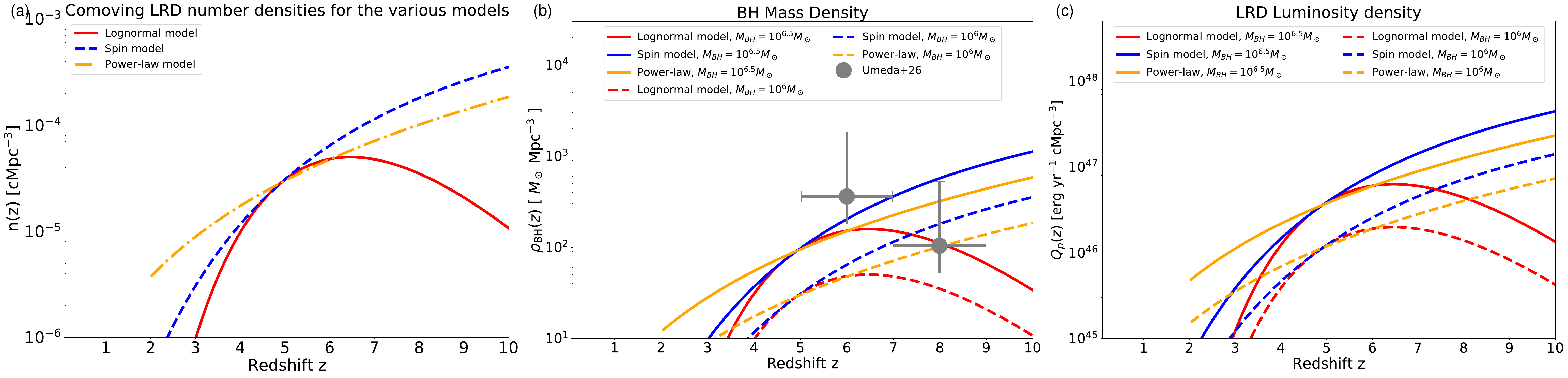}
    \caption{Comparison of the redshift evolution of the key quantities in the LRD scenario. (a) Comoving number densities of LRDs as a function of redshift for different population models. The line colors and styles are the same as in Fig.~\ref{fig:dFdz}. (b) BH mass density predicted by the same models. Line styles indicate the BH mass, with solid lines for $10^{6.5} M_\odot$ and dashed lines for $10^6 M_\odot$. Gray points are taken from \citet{Umeda_2026} and represent the BH mass density inferred from spectral fits of the BH-envelope model. (c) Proton luminosity density relevant for neutrino production, plotted in the same manner as panel (b). These panels outline the relation between LRD abundance, BH growth, and neutrino-related energetics.}
    \label{fig:number_z}
\end{figure*}

In this appendix, we summarize the three prescriptions for the comoving number density of LRDs used in Sections~\ref{sec:diffuse_convert} and \ref{sec:diffuse_neu}. Panel (a) of Fig.~\ref{fig:number_z} presents the corresponding redshift dependence for the different population models. As noted in Section~\ref{sec:diffuse_neu}, the conservative \textit{Lognormal model} peaks at $z \simeq 5$--7 and yields few sources at high redshift. On the other hand, the \textit{Spin model} predicts more abundant high-$z$ LRDs but exhibits a sharp decline toward low redshift. The \textit{Power-law model} follows a similar trend to the \textit{Spin model} at high redshift, while presenting high LRD abundances at low redshift.

Panels (b) and (c) of Fig.~\ref{fig:number_z} show the BH mass density (BHMD) and proton luminosity density for the same population models. The line colors indicate the model choice, and the line styles indicate the assumed BH mass, enabling comparison across the different prescriptions. We estimate the quantities in Panels (b) and (c) as $\rho_{\rm BH}(z) = M_{\rm BH}n(z)$ and $Q_{p}(z) = \epsilon_{p}L_{j}n(z)$. For the BHMD (see Panel (b)), all models lie within the observationally inferred range at $z \simeq 5$–7. At higher redshift ($z \simeq 7$–9), the \textit{Spin model} with $M_{\rm BH}=10^{6.5}M_\odot$ appears to exceed the inferred BHMD estimate \citep{Umeda_2026}. Nevertheless, the excess remains within the order-of-unity uncertainties, and we consider this deviation not to be in significant tension with current constraints. Panel (c) illustrates the corresponding evolution of the proton luminosity density, which reflects the energetics relevant for neutrino production and makes the connection between the population assumptions and their physical implications.

We also examine whether the adopted LRD population models are consistent with the energy budget of the cosmic infrared background (CIB). As the reprocessed optical emission from LRDs redshifts into the observed infrared bands, the integrated LRD population will contribute to the CIB. This contribution should not exceed the observed CIB intensity, and we compare the predicted LRD background with the measured CIB level. Adopting the maximal LRD abundance at $z\sim8$–$10$ and assuming that the bolometric luminosity is given by the Eddington luminosity of a $M_{\rm BH}\simeq10^{6.5}M_\odot$ BH ($L_{\rm bol}\approx L_{\rm Edd}$), the resulting background intensity is
approximately
\begin{eqnarray}
I_{\rm IR, LRD} &\approx& \left.\frac{c}{4\pi}\frac{n(z) L_{\rm bol}\Delta z}{H(z)(1+z)^2}\right|_{z=9} \nonumber \\
&\simeq& 5.2\times10^{-9}~{\rm erg~cm^{-2}~s^{-1}~sr^{-1}}\nonumber \\
&\times& L_{\rm bol,44.6}n_{-3.5} \left(\frac{\Delta z}{2.0}\right),
\label{eq:I_IR}
\end{eqnarray}
where $n_{-3.5} = n(z)/(10^{-3.5}~\rm Mpc^{-3})$ (see Panel (a) of Fig.~\ref{fig:number_z}). Eq.~(\ref{eq:I_IR}) shows that the expected infrared intensity is about three orders of magnitude below the observed CIB intensity of $I_{\rm CIB}\sim 10^{-5}~{\rm erg~cm^{-2}~s^{-1}~sr^{-1}}$ \citep{2013A&A...550A...4HESS, 2019PhRvD..99b3002Kalashev, 2019MNRAS.486.4233MAGIC}. Therefore, the LRD population satisfies both the BHMD and CIB energy--budget constraints even under optimistic assumptions.

Once the energy--budget constraints are verified, we proceed to describe the population models. For the \textit{Lognormal model}, \citet{2025ApJ...988L..22I} assumed that the activity time of LRDs, $t$, follows a lognormal probability distribution function,
\begin{eqnarray}
p(t)=\frac{1}{\sqrt{2\pi}\sigma_0t}\exp\left[-\frac{\{\ln(t/t_0)\}^2}{2\sigma_0^2}\right],
\end{eqnarray}
where $t_0\simeq 837 ~\rm Myr$ and $\sigma_0\simeq 0.327$ are the median and logarithmic width of the activity-time distribution, respectively. This lognormal form in cosmic time leads to a lognormal evolution in $\ln(1+z)$ for the comoving abundance, which is written as
\begin{eqnarray}
n_{\rm LN}(z)=\phi_0 f(z) \exp\left[-\frac{\{\ln(1+z)-\mu_z\}^2}{2\sigma_z^2}\right],
\label{eq:LN}
\end{eqnarray}
where $\log(\phi_0/{\rm Mpc^{-3}}) \simeq -5.2762 $ sets the overall normalization, and $\mu_z = \ln(1+z_0)$ with $z_0 = 6.53$ corresponding to the $t_0$, and $\sigma_z = 2\sigma_0/3 \simeq 0.218$ describe the peak redshift and the logarithmic width in $\ln(1+z)$, respectively. The remaining redshift dependence originates from the cosmic time-redshift relation and from the comoving volume element. These effects are contained in the function $f(z)$. With the high-redshift approximation $t(z)\approx 2/(3H_0\sqrt{\Omega_m})(1+z)^{-3/2}$ and the corresponding expressions for $ dt/dz$ and $dV_c/dz$, the correction function $f(z)$ can be written
\begin{eqnarray}
f(z)=\frac{(1+z)^{3/2}} {\left[s(1+z)^{1/2}-1\right]^{2}},
\end{eqnarray}
with $s=0.903$ calibrated for $\Omega_m = 0.307$ \citep{2025ApJ...988L..22I}.

For the \textit{Spin model}, \citet{2025ApJ...989L..19P} considered that the halo spin parameter $\lambda$ follows a lognormal distribution,
\begin{eqnarray}
p(\lambda)=\frac{1}{\lambda\sqrt{2\pi}\sigma_{\ln\lambda}}\exp\left[-\frac{\{\ln(\lambda/\bar\lambda)\}^2}{2\sigma_{\ln\lambda}^2}\right],
\end{eqnarray}
where $\bar\lambda$ is the median spin parameter and $\sigma_{\ln\lambda}$ its logarithmic width. An LRD can be formed when $\lambda<\lambda_{\rm LRD}(z)$. The critical value is given by $\lambda_{\rm LRD}(z)=\sqrt{2}\,{R_{\rm eff}}/{r_{200}(z)}$ and $r_{200}(z)=\left[{GM_{\rm halo}}/{(100H^2(z))}\right]^{1/3},$ where  $R_{\rm eff}$ is the threshold effective radius adopted to define compact LRDs and $M_{\rm halo}$ is the host halo mass. Here $r_{200}(z)$ denotes the virial radius of the dark matter halo, defined as the radius within which the mean density is 200 times the critical density of the Universe at redshift $z$. The fraction of eligible halos is
\begin{eqnarray}
f_{\rm LRD}(z)=\int_0^{\lambda_{\rm LRD}(z)}p(\lambda){\rm d}\lambda,
\end{eqnarray}
and the resulting comoving number density is obtained by
\begin{eqnarray}
n_{\rm spin}(z)=\phi_{\rm LBG}f_{\rm LRD}(z).
\label{eq:PL25}
\end{eqnarray}
Here, we adopt $\bar\lambda=0.05$, $\sigma_{\ln\lambda}=0.5$, and $M_{\rm halo}=10^{11} M_\odot$, which represent typical values inferred from simulations and observations. We set $R_{\rm eff} = 280~{\rm pc}$ to match the other models at $z\simeq 5.0 $, and this threshold lies within the observed size distribution of high-$z$ LRDs, which have median effective radii of $\sim150~{\rm pc}$ with a range of $\sim80$--$300~{\rm pc}$ \citep{Baggen2023ApJ}. The parameter $\phi_{\rm LBG}= 2.2\times10^{-3}~{\rm cMpc^{-3}~mag^{-1}}$ corresponds to the number density of Lyman break galaxies at $M_{\rm UV}\simeq-19$ \citep{Bouwens2021AJ}.

We also consider the \textit{Power-law model}, which takes the form of a simple power law and is normalized at $z=6$,
\begin{eqnarray}
n_{\rm PL}(z)=\phi_{z=5}\left(\frac{1+z}{1+z_0}\right)^{3},
\label{eq:PL}
\end{eqnarray}
where $z_0=5$ specifies the normalization redshift. The normalization is set to $\phi_{z=5}=3.0\times10^{-5}~{\rm cMpc^{-3}}$, chosen to match the order of magnitude inferred from Eqs.~(\ref{eq:LN}) and (\ref{eq:PL25}) at $z\simeq5$. This parametrization is not intended to represent a physically motivated model, but rather to provide a simple and transparent comparison with other astrophysical source classes.

\section{Secondary Pair Injection and Its Impact on the Dissipation Region} \label{app:secondary_inj}
As discussed in Section~\ref{sec:neu_esc_&_sec_inj}, $e^\pm$ pairs are inevitably produced either by the Breit–Wheeler process of photomeson-induced gamma rays or directly by the Bethe--Heitler process. In this Appendix, we estimate the number density of $e^\pm$ pairs produced through these channels, and compare it with the pre-existing electron number density to examine whether the dissipation region is radiation-mediated.

Photomeson production generates high-energy gamma rays that may trigger pair cascades in the form $\gamma^{(p\gamma)} \rightarrow \gamma^{(p\gamma)}\gamma^{(\rm thml)} \rightarrow e^\pm \xrightarrow{\rm IC} \gamma^{(\rm IC)} \rightarrow \cdots$ in the thermal photon bath of the surrounding accretion flows. The typical photon energy in the comoving frame is $\varepsilon'_\gamma \approx 2.8\Gamma_j k_B T_{\rm dis} \simeq 81~{\rm eV}~r_{\rm dis, 16}^{-1}M_{\rm BH,6.5}^{1/4}M_{\rm env, 6.5}^{1/4}$ $(\Gamma_j/2.0) \left(\ln(r_{\rm ph}/r_0)/10.3 \right)^{-1/4}$. An electron with Lorentz factor $\gamma_e$ upscatters such photons via IC scattering, producing secondary photons with $\varepsilon_{\gamma, \rm IC}' \approx \gamma_e^2\varepsilon_\gamma'$. For the cascade to continue, these IC photons must be able to pair-produce on the thermal photons through the Breit--Wheeler process. Once the upscattered photon energy drops below the pair-creation threshold, the cascade stalls. A conservative estimate of this condition is obtained by requiring $\varepsilon_{\gamma, \rm IC}' \lesssim m_e c^2$, where $m_e$ is the electron mass. This yields a minimum electron Lorentz factor
\begin{eqnarray}
{\gamma'}_{e,\min}^{(\gamma\gamma)} &\approx& \left(\frac{m_e c^2}{2.8 \Gamma_j k_B T_{\rm dis}}\right)^{1/2} \nonumber \\ 
&\simeq& 79~r_{\rm dis, 16}^{1/2}M_{\rm BH,6.5}^{-1/8}M_{\rm env,6.5}^{-1/8} \!\left(\frac{\Gamma_j}{2.0}\right)^{-1/2} \nonumber \\
&\times& \left(\frac{\ln(r_{\rm ph}/{r_{0}})}{10.3}\right)^{1/8}.
\end{eqnarray}
Electrons injected at high energies rapidly cool via IC scattering down to $\gamma_{e,\min}^{(\gamma\gamma)}$, but cannot sustain further Breit--Wheeler process below this point. Consequently, the cascade self-quenches after a few generations.

The gamma-ray luminosity associated with the photomeson production can be estimated from the fraction of the proton power above the photomeson threshold, $\varepsilon'_p > \varepsilon^{' \rm eff}_{{\rm thr},p\gamma}$. Approximately, half of this power is channeled into neutral pions, which subsequently decay into gamma rays, while the other half goes into charged pions. Thus, the gamma-ray luminosity is approximated to be $L'_{\gamma,p\gamma}\approx 0.5\epsilon_p L'_j \ln(\varepsilon'_{p,\max}/\varepsilon^{' \rm eff}_{{\rm thr},p\gamma}) / \ln(\varepsilon'_{p,\max}/\varepsilon'_{p,\min}) \simeq 3.2\times10^{41}~{\rm erg~s^{-1}}L_{j,44.6}\epsilon_{p,-1} (\Gamma_j/2.0)^{-2}\!\left(\ln (\varepsilon'_{p,\max}/\varepsilon^{' \rm eff}_{{\rm thr},p\gamma})/0.9\right)$ $({\ln (\varepsilon'_{p,\max} / \varepsilon'_{p,\min})}/15)^{-1}$, and the number density of secondary pairs relative to the original electron density is
\begin{align}
\frac{n'_{\pm,p\gamma}}{n_j'} &\approx \frac{L'_{\gamma, p\gamma}}{L'_j}\frac{m_p}{\gamma'^{(\gamma\gamma)}_{e, \rm min}m_e} \nonumber \\
&\simeq 7.2\times10^{-2} ~\epsilon_{p,-1}
r_{\rm dis, 16}^{-1/2}M_{\rm BH,6.5}^{1/8}M_{\rm env,6.5}^{1/8} 
\left(\frac{\Gamma_j}{2.0}\right)^{1/2} \nonumber \\
&\times \left(\frac{\ln(r_{\rm ph}/{r_{0}})}{10.3}\right)^{-1/8}
\left(\frac{\ln(\varepsilon'_{p,\max}/\varepsilon^{' \rm eff}_{{\rm thr},p\gamma})}{0.9}\right) \nonumber \\
&\times \left(\frac{\ln(\varepsilon'_{p,\max}/\varepsilon'_{p,\min})}{15}\right)^{-1}
\end{align} 
This scaling shows that the additional Thomson optical depth from secondary pairs is $\tau_{e^\pm}\approx n'_{\pm,p\gamma}\sigma_{\rm T} r_{\rm dis}/\Gamma_j < 1$, and thus, the dissipation region remains optically thin and not radiation-mediated.

We next estimate the number density of $e^\pm$ pairs produced via the Bethe--Heitler process. The threshold condition gives minimum Lorentz factor of the produced $e^\pm$ pairs as
\begin{eqnarray}
{\gamma'}_{e,\min}^{(\rm BH)} &\approx& \frac{m_e c^2}{2.8 \Gamma_j k_B T_{\rm dis}} \nonumber \\
&\simeq& 6.3\times10^3 ~r_{\rm dis, 16}M_{\rm BH,6.5}^{-1/4}M_{\rm env,6.5}^{-1/4} \left(\frac{\Gamma_j}{2.0}\right)^{-1}\nonumber \\
&\times&\left(\frac{\ln(r_{\rm ph}/{r_{0}})}{10.3}\right)^{1/4}.
\end{eqnarray}
It should be noted that protons with $\varepsilon'_p > \varepsilon^{' \rm eff}_{{\rm thr},p\gamma}$ preferentially lose their energy via photomeson production rather than the Bethe--Heitler process (see Fig.~\ref{fig:timescale}). Therefore, the effective energy range that contributes to Bethe--Heitler pair production is limited to $\varepsilon'_{\rm thr,BH} < \varepsilon'_p < \varepsilon^{' \rm eff}_{{\rm thr},p\gamma}$, where $\varepsilon'_{\rm thr,BH} ~=\gamma'^{(\rm BH)}_{e,\rm min}\varepsilon'_{p,\min}$ is the threshold proton energy for the Bethe--Heitler process. All of this energy loss proceeds directly into $e^\pm$ pairs, and thus the pair injection luminosity is given by $L'_{e^\pm,\rm BH}\approx \epsilon_p L'_j \ln(\varepsilon^{' \rm eff}_{{\rm thr},p\gamma}/\varepsilon'_{{\rm thr, BH}}) / \ln(\varepsilon'_{p,\max}/\varepsilon'_{p,\min})$. The number density of secondary pairs relative to the original electron density is estimated to be
\begin{align}
\frac{n'_{\pm,{\rm BH}}}{n_j'} &\approx \frac{L'_{e^\pm, {\rm BH}}}{L'_j}\frac{m_p}{\gamma'^{(\rm BH)}_{e, \rm min}m_e} \nonumber \\
&\simeq 9.9\times10^{-3}~ 
\epsilon_{p,-1}r_{\rm dis, 16}^{-1}M_{\rm BH,6.5}^{1/4}M_{\rm env,6.5}^{1/4} 
\left(\frac{\Gamma_j}{2.0}\right) \nonumber \\
&\times  \left(\frac{\ln(r_{\rm ph}/{r_{0}})}{10.3}\right)^{-1/4}
\left(\frac{\ln(\varepsilon^{' \rm eff}_{\rm thr,p\gamma}/\varepsilon'_{\rm thr,BH})}{4.9}\right) \nonumber \\
&\times \left(\frac{\ln(\varepsilon'_{p,\max}/\varepsilon'_{p,\min})}{15}\right)^{-1},
\end{align}
which is lower than in the photomeson case, and the associated Thomson optical depth is lower than unity. Thus, the Bethe--Heitler process cannot render the dissipation region radiation-mediated. Even if we take the target photons in the Wien tail, the change in $\gamma'^{(\rm BH)}_{e,~\rm min}$ is only a factor of a few. Hence, the estimates for $n'_{\pm,{\rm BH}}/n'_j$ and the Thomson optical depth vary by at most the same factor, and the conclusion above remains unchanged.

In either case, through cascades initiated by photomeson-induced gamma rays or through the Bethe--Heitler process, the number of produced secondary pairs remains below that of pre-existing electrons in the dissipation region. Thus, the additional Thomson optical depth is negligible, and the dissipation region is not radiation-mediated within our parameter range.

\bibliographystyle{apsrev4-2}
\bibliography{sample631}



\end{document}